\documentclass[
    aps,
    prd,
    reprint, % Use 'preprint' for single-column/double-spaced draft
    groupedaddress,
    nofootinbib,
    amsmath,
    amssymb,
    amsfonts
]{revtex4-2}
\def\dd{{\rm d}}

\usepackage{graphicx}  % Include figure files
\usepackage{enumitem}
\usepackage{orcidlink}
\usepackage{dcolumn}   % Align table columns on decimal pointhttp://arxiv.org/abs/2608.02798
\usepackage{bm}        % Bold math
\usepackage{hyperref}  % Hyperlinks

\hypersetup{
    colorlinks=true,
    linkcolor=red,   % \ref and \eqref (also sections, figures, tables)
    citecolor=blue,  % \cite
    urlcolor=blue,
}

\DeclareMathAlphabet{\mathdutchcal}{U}{dutchcal}{m}{n}
\SetMathAlphabet{\mathdutchcal}{bold}{U}{dutchcal}{b}{n}

\usepackage[dvipsnames]{xcolor}

\begin{document}

% Title of the paper
\title{Asymmetric Quantum Oppenheimer-Snyder Collapse}

% Authors and Affiliations
\author{Micha{\l} Bobula\orcidlink{0000-0002-4928-9568}}
\email{michal.bobula@uwr.edu.pl}
\affiliation{University of Wroc{\l}aw, Faculty of Physics and Astronomy,
Institute of Theoretical Physics, Plac Maksa Borna 9, 50-204 Wroc{\l}aw, Poland}

\author{Tomasz Paw{\l}owski\orcidlink{0000-0002-5592-4804}}
\email{tomasz.pawlowski@uwr.edu.pl}
\affiliation{University of Wroc{\l}aw, Faculty of Physics and Astronomy,
Institute of Theoretical Physics, Plac Maksa Borna 9, 50-204 Wroc{\l}aw, Poland}

% Date
% Abstract
\begin{abstract}
We study black hole formation resulting from the collapse of a homogeneous and isotropic dust ball within a framework built upon loop quantum cosmology. Using loop dynamics formulated for Lema\^{\i}tre-Tolman-Bondi spacetimes—which reduce to the asymmetric loop quantum cosmological bounce scenario—we analyze the dust ball undergoing power-law contraction followed by a de Sitter-like expansion. We discover an emergence of a physical shock at the dust ball surface, characterized by a continuous induced metric and a discontinuous extrinsic curvature arising from a dynamical mismatch between the expanding interior and the contracting exterior vacuum shells. Furthermore, we explicitly derive two variants of Schwarzschild-like line elements for the vacuum sector, and we demonstrate that the resulting spacetime is geodesically complete featuring a rich causal structure. This establishes, for the first time, fully dynamical regular black hole formation within a framework consistent with loop quantum cosmology.
\end{abstract}

\maketitle

\section{Introduction}

Since the emergence of singularities signals the outright breakdown of classical predictability, resolving them remains the primary imperative for extending general relativity (GR). However, any fundamental successor theory must achieve this without sacrificing theoretical integrity. It is not enough to resolve the Big Bang singularity or tame the interior of a black hole independently: a viable quantum or effective theory of gravity must yield a self-consistent description that unifies both cosmological dynamics and the physics of extreme gravitational collapse. 

In this context, the dynamical black hole formation framework of the Oppenheimer-Snyder (OS) collapse scenario becomes critical testing field, both in its classical formulation \cite{Oppenheimer:1939ue, Datt:1938uwc, Markovic:1999di} and when extended to effective theories \cite{Peleg:1995gg, Govender:2001gq, Bojowald:2008ja, Bambi:2013caa, Zhang:2014bea, Schmitz:2019jct, Piechocki:2020bfo, Schmitz:2020vdr, Munch:2020czs, BenAchour:2020bdt, BenAchour:2020mgu, BenAchour:2020gon, Bonanno:2023rzk, Kelly:2020lec, Kelly:2020uwj, Husain:2021ojz, Husain:2022gwp, Giesel:2022rxi, Lewandowski:2022zce, Bobula:2023kbo, Fazzini:2023scu, Han:2023wxg, Bobula:2024jlh, Bobula:2024ywp, Bobula:2024chr, Cafaro:2024vrw, Cafaro:2026twe, Duque:2023syb, Bojowald:2024ium, Shi:2024vki, Ou:2025bbv, Alonso-Bardaji:2023qgu, Malafarina:2022oka, Shojai:2022pdq, Bueno:2025gjg, Li:2025kou}. This framework unifies these seemingly disparate physical problems into a single scenario. Due to the inherent symmetries of the problem, the collapse of a homogeneous and isotropic dust ball is governed by cosmological dynamics in its interior, while the exterior vacuum is described by a spherically symmetric black hole geometry. Consequently, extending the OS collapse scenario beyond classical GR provides a rigorous benchmark for evaluating the self-consistency—and constraining the parameter space—of any proposed successor theory.

More recently, several quantum-corrected OS collapse scenarios have garnered considerable attention in the literature. These models—rooted in or inspired by leading quantum gravity candidates like loop quantum gravity (LQG) \cite{Husain:2021ojz, Husain:2022gwp, Lewandowski:2022zce}, asymptotically safe gravity \cite{Bonanno:2023rzk}, and string theory \cite{Bueno:2025gjg}—share a crucial feature: the collapsing dust ball avoids the classical Schwarzschild-like singularity. Instead, the collapse typically follows one of two primary trajectories: a high-energy bounce \cite{Bambi:2013caa, Zhang:2014bea, Schmitz:2019jct, Piechocki:2020bfo, Schmitz:2020vdr, Munch:2020czs, BenAchour:2020bdt, BenAchour:2020mgu, BenAchour:2020gon, Kelly:2020lec, Kelly:2020uwj, Husain:2021ojz, Husain:2022gwp, Giesel:2022rxi, Lewandowski:2022zce, Bobula:2023kbo, Fazzini:2023scu, Han:2023wxg, Bobula:2024ywp, Bobula:2024chr, Cafaro:2024vrw, Cafaro:2026twe, Duque:2023syb, Bojowald:2024ium, Shi:2024vki, Ou:2025bbv, Alonso-Bardaji:2023qgu, Bueno:2025gjg}, or a dramatic deceleration leading to a so-called \emph{long-squeeze} indefinite collapse \cite{Malafarina:2022oka, Bonanno:2023rzk, Bobula:2024jlh, Li:2025kou}. This dichotomy closely mirrors a familiar parallel in cosmology, where either a bouncing or an inflationary regime is conventionally proposed to resolve the fundamental puzzles of the early universe.

By demanding consistency between interior dynamics and exterior geometry, these models represent a major conceptual advance over previous literature. Historically, researchers have tended to decouple these domains, proposing singularity-free black hole geometries without compatible collapse dynamics, or, conversely, exploring cosmological resolutions devoid of a corresponding black hole exterior. Because the modern OS constructions unify these regimes, they provide a far more complete foundation for theory-building. This evolution is particularly evident when re-examining older black hole constructions within midisuperspace LQG \cite{Ashtekar:2005qt, Gambini:2008dy, Gambini:2013ooa, Gambini:2013hna, Olmedo:2017lvt, Ashtekar:2018lag, Ashtekar:2018cay, Bodendorfer:2019cyv}, which largely focused on static geometries rather than the full dynamical collapse.

In this work, we focus on constructions derived from symmetry-reduced LQG dynamics \cite{ Bojowald:2008ja, Munch:2020czs, BenAchour:2020bdt, BenAchour:2020mgu, BenAchour:2020gon, Kelly:2020lec, Kelly:2020uwj, Husain:2021ojz, Husain:2022gwp, Giesel:2022rxi, Lewandowski:2022zce, Bobula:2023kbo, Fazzini:2023scu, Han:2023wxg, Bobula:2024ywp, Bobula:2024chr, Cafaro:2024vrw, Cafaro:2026twe, Duque:2023syb, Bojowald:2024ium, Shi:2024vki, Ou:2025bbv, Alonso-Bardaji:2023qgu}. Recently, the OS collapse scenario has been extensively studied within the framework of Loop Quantum Cosmology (LQC) \cite{Ashtekar:2011ni, Agullo:2023rqq}—a cosmological minisuperspace reduction of LQG. Mainstream LQC models are well known for resolving the Big Bang singularity via a (typically time-symmetric) quantum bounce \cite{Ashtekar:2006rx, Ashtekar:2006uz, Ashtekar:2006wn}. When applied to the OS scenario, this mechanism yields a gravitational collapse model in which the dust ball bounces in the deep Planckian regime and reemerges into a new asymptotic region \cite{Lewandowski:2022zce, Bobula:2023kbo, Fazzini:2023scu}. This produces a global causal structure resembling that of a Reissner-Nordström spacetime. However, while the classical central singularity is successfully resolved, this framework suffers from notable shortcomings. Most prominently, the analytic extension of the vacuum exterior unavoidably generates a Reissner-Nordström-like timelike singularity \cite{Bobula:2023kbo, Lewandowski:2022zce, Bobula:2024chr}.

Despite its successes, mainstream LQC remains an incomplete description of the early universe. Because it fails to produce inflation purely from quantum geometry, LQC is forced to rely on the ad hoc insertion of a scalar inflaton field—a common but artificial fix in cosmological models. To address this, recent efforts have focused on refining LQC dynamics to better reflect the underlying kinematics of full LQG. A key breakthrough in this program is the asymmetric quantum cosmological bounce \cite{Assanioussi:2018hee, Assanioussi:2019iye}, derived via alternative quantization schemes for the gravitational Hamiltonian constraint. When employing either a scalar field or dust as a relational clock, this refined construction yields a profound modification to the universe's evolution: a de Sitter-like contraction phase that evolves through the quantum bounce directly into a power-law expansion. The natural question we answer in this work is the following: what is the exact prediction of this type of dynamics when implemented for a quantum-corrected OS collapse scenario.\footnote{Although preliminary results addressing this question were presented in \cite{Ou:2025bbv}, which studied quasi-normal modes for the vacuum solution derived via surface matching, the fully dynamical physical picture was not analysed.}

Recently, both symmetric and asymmetric effective LQC-like dynamics have been generalized to Lema\^{i}tre-Tolman-Bondi (LTB) spacetimes \cite{Giesel:2022rxi, Giesel:2023hys, Giesel:2023tsj, Giesel:2024mps, Fazzini:2026uqp, Giesel:2026pjj}, establishing a highly suitable basis for studying the OS collapse scenario. In this work, we present a detailed analysis of the effective dynamical equations governing the asymmetric dynamics derived in \cite{Giesel:2024mps}. By applying these equations to the OS collapse model, we explicitly derive the exact geometric evolution and the resulting global causal structure. For the first time, we provide a fully dynamical formation of a regular black hole within a framework based on LQC, demonstrating the geodesic completeness of the resulting spacetime.

A notable consequence of the studied dynamics is the unexpected emergence of a physical \emph{shock} in the post-bounce region. We define this shock as a spacetime hypersurface across which the gravitational field exhibits a distinct discontinuity. In general, such a shock can manifest in two ways: a strong discontinuity affecting both the induced metric and extrinsic curvature, or a weaker discontinuity in which the extrinsic curvature jumps while the induced metric maintains $C^0$ continuity. Although a strong shock was previously conjectured to occur in the symmetric LQC-based OS collapse scenario \cite{Husain:2021ojz, Husain:2022gwp}, subsequent investigations revealed it to be a coordinate artifact \cite{Fazzini:2023scu}. Recent work demonstrates that physical shocks in symmetric bounce models are restricted to inhomogeneous collapse scenarios \cite{Husain:2021ojz, Husain:2022gwp, Bobula:2024chr, Fazzini:2023ova, Liu:2025fil} (beyond the strictly homogeneous OS model) and are characterized solely by discontinuous extrinsic curvature \cite{Bobula:2026zlq}, where the shock is tightly linked to shell-crossing singularities. Notably, our present work establishes the existence of a weak $C^0$ shock inherent to the asymmetric OS collapse model—a key feature distinguishing it from the symmetric case, in which such shocks are absent.

This work extends beyond current regular black hole formation models within the literature \cite{Bonanno:2023rzk, Bobula:2024jlh, Bueno:2024eig, Bueno:2025gjg, Bueno:2025zaj, Li:2025kou} by offering a rigorous analysis of unexplored fully dynamical asymmetric bounce. We argue that our conclusions—particularly those concerning the emergence of shocks—are not limited to frameworks based on LQC but possess a broader universality potentially applicable to any gravitational collapse model characterized by an asymmetric bounce.

The paper is organized as follows. Sec. \ref{sec:asymmetric_review} provides a review of the asymmetric effective LTB dynamics. In Sec. \ref{sec:analysis_of_asymmetric}, we investigate the geometry of the asymmetric OS collapse scenario, focusing on shock formation and the resulting global causal structure. Sec. \ref{sec:vacuum} details the corresponding vacuum solutions governed by these dynamics. Finally, we summarize our results and provide concluding remarks in Sec. \ref{sec:discussion}.

%%%%%%%%%%%%%%%%%%%%%%%%%%%%%%%%%%%%%%%%%%%%%%%%%%%%%%%%%%%%%%%%%%

%%%%%%%%%%%%%%%%%%%%%%%%%%%%%%%%%%%%%%%%%%%%%%%%%%%%%%%%%%%%%%%%%%
\section{Asymmetric loop quantum LTB dynamics} \label{sec:asymmetric_review}

We outline the asymmetric LQC framework \cite{Assanioussi:2018hee, Assanioussi:2019iye} generalized to LTB spacetimes \cite{Giesel:2024mps, Fazzini:2026uqp, Giesel:2026pjj} and present its effective dynamics. The classical system prior to quantization consists of an irrotational dust field coupled to gravity, described by the action
\begin{equation}
    S= \frac{1}{16 \pi G} \int \mathrm{d} \mathdutchcal{y} \sqrt{-\mathrm{det(g)}} \left( \mathcal{R}-\frac{\rho}{2}\left[g^{\mu \nu} T_{;\mu} T_{;\nu}+1\right] \right) \,,
\end{equation}
where $\mathcal{R}$ is the four-dimensional Ricci scalar, $g_{\mu\nu}$ is the spacetime metric, and $T$ is the dust field with energy density $\rho$. The effective, Ehrenfest-like semiclassical equations of quantum dynamics—where quantization proceeds after symmetry reduction—were derived using a 3+1 decomposition of the spacetime metric under spherical symmetry and LTB-like gauge choices in \cite{Giesel:2024mps, Giesel:2026pjj, Fazzini:2026uqp}. These equations are compatible with the asymmetric LQC bounce scenario when reduced to a homogeneous and isotropic matter distribution. A more popular treatment---mainstream LQC simplifies the Hamiltonian constraint by splitting the Lorentzian term into part proportional to Euclidean and a $3$-dimensional Ricci scalar which subsequently is set to zero as it classically vanishes in the spatially flat cosmological spacetimes under consideration. This leads to a standard, symmetric $\rho^2$ correction to the effective Friedmann equation. On the other hand, the framework used in this work keeps the strict adherence to Thiemann’s regularization scheme \cite{Thiemann:1996av, Thiemann:1996aw}. The resulting effective dynamics introduces then higher-than-quadratic corrections in the energy density $\rho$, fundamentally breaking the time-reversal symmetry between the contracting and expanding phases.

Within this framework, the effective line element for the marginally bound collapse scenario is given by
\begin{equation} \label{met}
    \mathrm{d}\mathdutchcal{s}^2 = -\mathrm{d}T^2+\left(\frac{\partial R(T,x)}{\partial x} \right)^2 \mathrm{d}x^2  +   R(T,x)^2 \mathrm{d}\Omega^2 \,,
\end{equation}
where $\mathrm{d}\Omega^2 := \mathrm{d}\theta^2 + \sin^2\theta \, \mathrm{d}\phi^2$ is the metric on the unit two-sphere. The use of $T$ as the time coordinate reflects the dust time gauge—meaning the value of the dust field serves as a physical clock for the system. This formulation is consistent with an LTB-like coordinate frame, in which the time coordinate corresponds to the proper time of observers comoving with the dust. The effective Friedmann equation is \cite{Giesel:2024mps, Fazzini:2026uqp, Giesel:2026pjj}
\begin{equation} \label{friedmann}
\frac{\dot{R}(t, x)}{R(t, x) }= \pm \frac{\sqrt{\frac{1}{2}-\gamma^2 \mathcal{S}} \sqrt{\mathcal{S} \pm \sqrt{1-2 \gamma^2 \mathcal{S}}+1}}{\alpha_{\Delta}\left(\gamma^2+1\right)}\,,
\end{equation}
where $\mathcal{S}:= 2 \alpha_{\Delta}^2\left(\gamma^2+1\right)\left(\frac{2 G M(x)}{R(t, x)^3}\right)$, $\gamma =0.23$ is the dimensionless Barbero-Immirzi parameter we are fixing accordingly to the results in \cite{Meissner:2004ju, Domagala:2004jt}, $\alpha_\Delta:=\gamma \sqrt{\Delta}$, and $\Delta := 4 \pi l_\mathrm{Pl}^2$ is the \emph{area gap}, expressed in terms of the Planck length $l_\mathrm{Pl}$, which constitutes the smallest non-zero eigenvalue of the area operator in LQG. Here, $M(x)$ is a conserved quantity interpreted as the accumulated dust mass at the comoving radial distance $x$—that is, the total mass enclosed within a dust shell at coordinate $x$. It is related to the energy density as follows
\begin{equation} \label{massdef}
M(x)=4 \pi \int_0^x \rho(T,\tilde{x}) R^2\left(\partial_{\tilde{x}} R\right) \mathrm{d} \tilde{x} \,,
\end{equation}
In the Friedmann equation \eqref{friedmann}, the $\pm$ signs are independent, yielding four distinct possibilities for the evolution in the vicinity of the chosen initial time slice: (i) power-law contraction, (ii) power-law expansion, (iii) de Sitter-like contraction, and (iv) de Sitter-like expansion. This variety contrasts with classical LTB dynamics—recovered in the limit $\alpha_\Delta \to 0$ taken along with fixing the minus sign inside the square root in \eqref{friedmann}—where only options (i) and (ii) remain viable. In this work, we focus on power-law contraction in the vicinity of the initial time slice, as this is the only viable option for modeling stellar gravitational collapse. The solutions of \eqref{friedmann} are given in $\eta$-parametric form \cite{Giesel:2024mps, Giesel:2026pjj, Fazzini:2026uqp}
\begin{equation} \label{Rsol}
\begin{aligned}
    R(T, x)=\sqrt[3]{\frac{  G M(x)\left(4 \alpha_{\Delta}^2 \gamma^2+9 \eta^2\right)^2}{18 \eta^2-8 \alpha_{\Delta}^2 \gamma^4}} & \,, \\ s(x)-T= \eta-\frac{2}{3} \alpha_{\Delta}\left(\gamma^2+1\right)  \tanh ^{-1} &\left(\frac{2 \alpha_{\Delta} \gamma^2}{3 \eta}\right) \,.
\end{aligned}
\end{equation}
where $\eta=\eta(T,x)$ and $s(x)$ is a function that fixes the coordinate freedom to redefine the comoving radius $x$ relative to the areal radius $R$. In this work, we focus on the collapse of a homogeneous and isotropic dust ball—the OS collapse scenario. Therefore, on the initial time slice $T=0$, we specify the initial energy density profile to be
\begin{equation} \label{rho0}
    \rho(0,x) = \begin{cases}
        \rho_0  &\text{for}\;x\leq x_b \,,\\
        0  &\text{for} \;\; x>x_b \,,
    \end{cases}
\end{equation}
along with the choice
\begin{equation} \label{R0}
    R(0,x) = x \,.
\end{equation}
These choices allows us to explicitly evaluate \eqref{massdef}, yielding the mass profile
\begin{equation} \label{Mx}
    M(x) = \begin{cases}
        \frac{4}{3} \pi  \rho_0 x^3 &\text{for}\;x\leq x_b \,, \\
        \frac{4}{3} \pi  \rho_0 x_b^3 &\text{for}\;x > x_b \,.
    \end{cases}
\end{equation}
Next, we define the total mass of the dust ball as the constant $M:=\frac{4}{3} \pi \rho_0 x_b^3$. This setup fully determines the function $s(x)$ in Eq. \eqref{Rsol}, as we demonstrate explicitly in Appendix \ref{appendix_sx}.

In Fig. \ref{fig_shells}, we plot the $R(T,x)$ solutions for the OS collapse scenario on the areal radius $R$ versus proper time $T$ plane, with each shell labeled by a constant $x$ coordinate. Two main families of shells originate from the initial time slice: (i) interior shells carrying dust, and (ii) vacuum (test) shells that contain no matter, tracking instead timelike geodesics in the external vacuum region.  Here, we also highlight a non-trivial feature shared by both symmetric and—as demonstrated in the next parts of this work—asymmetric bouncing LTB collapse models: the discontinuity of the proper time $T$ across the surface of the collapsing dust ball after the bounce. This phenomenon was originally uncovered for the symmetric LQC-based OS collapse in \cite{Fazzini:2023scu}, resolving previous claims regarding the presence of physical shocks in that scenario. Specifically, this proper time discontinuity had been misinterpreted as a fundamental spacetime discontinuity in the induced metrics and extrinsic curvatures. This misunderstanding was rooted in integration methods that implicitly assumed a continuous time coordinate across the dust ball surface post-bounce. In the asymmetric OS collapse model analyzed here, this time discontinuity manifests as follows. Crucially, in Fig. \ref{fig_shells}, the intersection of a vacuum shell with the dust ball surface is not the point where the shell physically enters the interior. Were that the case, the proper time $T$ would be continuous across the boundary. Instead, every vacuum shell in Fig. \ref{fig_shells} enters the interior in the vicinity of the horizontal pink line, more precisely at the intersection with the $\tilde{R}(\tilde{\mathcal{T}}, \tilde{\mathdutchcal{x}})$ dashed curve. The detailed analysis of this discontinuity is presented in Sec. \ref{sec:analysis_of_asymmetric}. As we elaborate further in this work, the key distinction between the symmetric \cite{Husain:2021ojz, Husain:2022gwp, Lewandowski:2022zce, Bobula:2023kbo, Fazzini:2023scu} and asymmetric OS collapse scenarios lies in the presence of a genuine physical shock alongside this proper time discontinuity—a situation fundamentally different from the symmetric OS collapse.

\begin{figure*}
    \centering
    \includegraphics[width=0.7\linewidth]{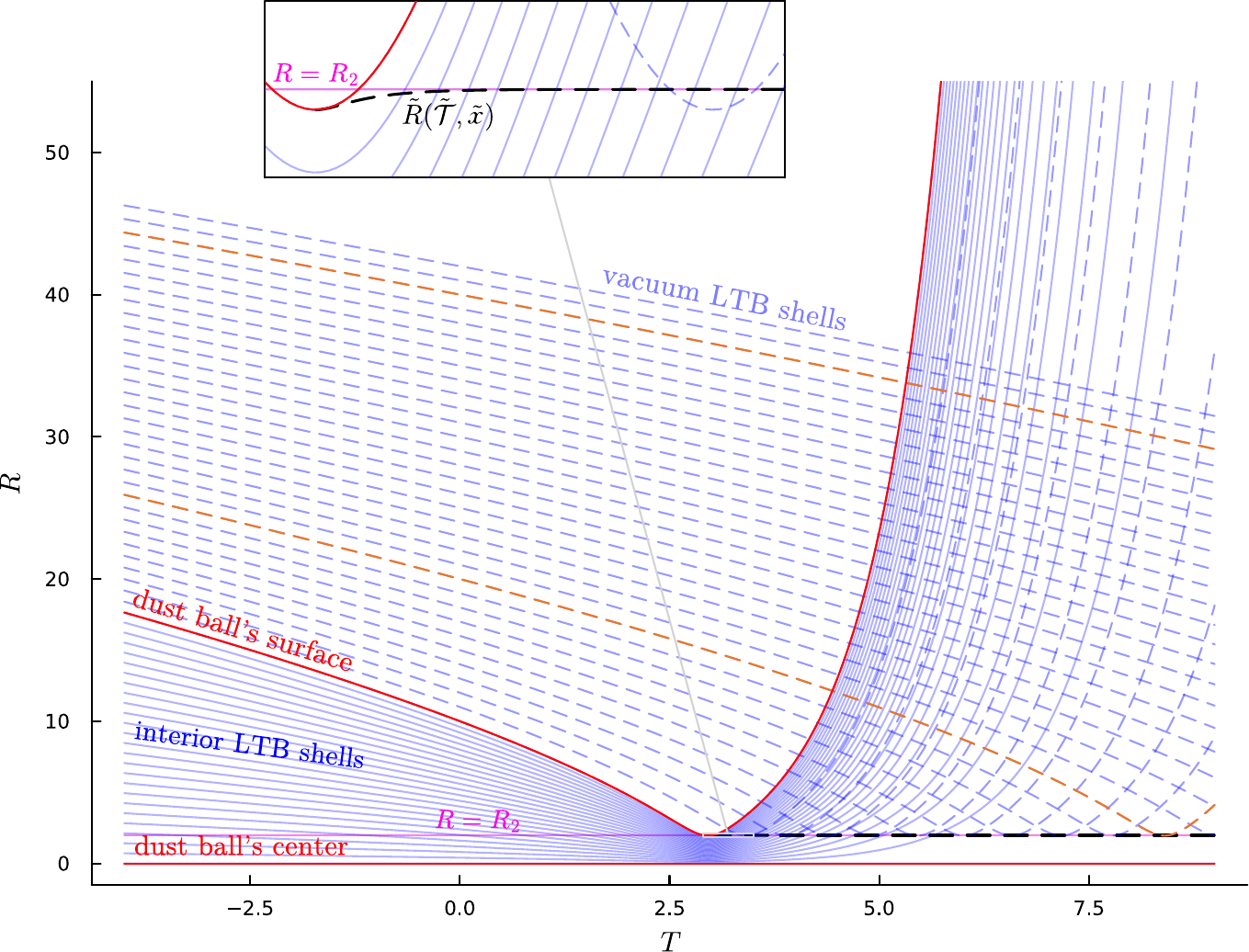}
    \caption{Timelike geodesics (or constant-$x$ shells) on the areal radius $R$ versus proper time $T$ plane, corresponding to the $R(T,x)$ solution for the asymmetric OS collapse scenario. The dust ball undergoes power-law contraction and subsequently expands exponentially after the bounce. For this plot, the parameters are chosen as $\rho_0=0.006$ and $x_b=10$ in Planck units. \emph{Inset.} The pink line of constant $R=R_2$ is plotted alongside the trajectory $\tilde{R}(\tilde{\mathcal{T}}(T,x_b), \tilde{\mathdutchcal{x}}(T,x_b) )$ (black dashed line). The intersection of the vacuum shell with this trajectory determines the time $\tilde{\mathcal{T}}$ (readable from the corresponding $T$-coordinate on the horizontal axis) at which the vacuum shell impacts the surface of the dust ball. By utilizing the blind-shooting algorithm, we have found the following initial conditions $T^\mathrm{bs}_1 \approx 2.98$, $\tilde{\mathdutchcal{x}}(T^\mathrm{bs}_1) \approx 10.12$ in Planck units. }  
    \label{fig_shells}
\end{figure*}

Analyzing the bounce condition $\partial_T R(T,x) = 0$ reveals that, for each shell, the bounce occurs at the parameter value $\eta = \eta_\mathrm{min}:= \frac{2}{3}\sqrt{\alpha_\Delta^2 \gamma^2 \left( 1+ 2 \gamma^2 \right)}$ and at the minimal radius $R=R_\mathrm{min}(x) := 2 \left[\alpha_\Delta^2 \gamma^2 \left( \gamma^2 +1 \right)  M(x) \right]^{1/3}$. As established in \cite{Giesel:2024mps}, the Kretschmann scalar—the quadratic contraction of the Riemann curvature tensor—diverges exactly at the bounce for each shell in the vacuum region. Conversely, we find that for each interior shell, this curvature scalar remains bounded. 

This divergence, occurring in particular at the surface of the collapsing dust ball -- that is, for the boundary shell at $x=x_b$, at the bounce time $T=T_\mathrm{bounce}$—indicates the presence of a physical shock (in symmetric OS collapse \cite{Husain:2021ojz, Husain:2022gwp, Lewandowski:2022zce, Bobula:2023kbo, Fazzini:2023scu} scenario there is no such divergence). Because the corresponding trajectory does not terminate there but instead possesses an unambiguous extension beyond that point, this behavior signals a weak singularity associated with the shock. To investigate its nature in greater detail, in the next section we employ junction conditions in order to determine the exact trajectory of the shock.

\label{sec:theory}
%%%%%%%%%%%%%%%%%%%%%%%%%%%%%%%%%%%%%%%%%%%%%%%%%%%%%%%%%%%%%%%%%%

%%%%%%%%%%%%%%%%%%%%%%%%%%%%%%%%%%%%%%%%%%%%%%%%%%%%%%%%%%%%%%%%%%
\section{Post-bounce shock in the asymmetric collapse scenario} \label{sec:analysis_of_asymmetric}

Our objective here is to understand the physical shock developing specifically at the surface of the collapsing dust ball at the bounce point, where the Kretschmann scalar diverges (see Sec. \ref{sec:asymmetric_review}). To do so, we must distinguish between different time parametrizations natural to the two spacetime regions distinguished in the previous section. While Fig. \ref{fig_shells} illustrates the $T$-time profile of the surface according to a comoving observer, it does not capture directly the timing from the perspective of the infalling vacuum shells, that is, the explicit $T$-times at which these exterior shells successively impact the dust ball's boundary. Here, we detail how to determine these times.

We will enforce a minimal junction condition requiring the continuity of the induced metric across the collapsing dust ball's surface. In Sec. \ref{secsub_junction}, we analyze this condition for two general LTB-like line elements. Subsequently, in Sec. \ref{secsub_detailed}, we apply these results to the asymmetric collapse scenario; this detailed study yields a unique $T$-parametrization for the vacuum shells impacting the boundary, thereby enabling us to determine the relevant geometric properties there.

\subsection{Junction of two LTB spacetimes} \label{secsub_junction}

Let us examine a general setup not necessarily confined to specific asymmetric OS collapse scenario considered so far. We investigate the detailed matching of two arbitrary LTB spacetimes in marginally bound configuration, where the matching hypersurface is formed out of (foliated by) geodesic trajectories in one of the spacetimes, while remaining a general trajectory (not necessarily a geodesic) in the other.

The to-be-matched LTB line elements are
\begin{equation} \label{met1_glu}
    \dd \mathdutchcal{s}^2 = -\dd T^2 + R_{,x} \dd x^2 +R^2 \dd\Omega^2 \,,
\end{equation}
and
\begin{equation} \label{met2_glu}
    \dd \tilde{\mathdutchcal{s}}^2 = -\dd \tilde{ \mathcal{T} }^2 + \tilde{R}_{,\tilde{\mathdutchcal{x}}} \dd \tilde{\mathdutchcal{x}}^2 +\tilde{R}^2 \dd \tilde{\Omega}^2 \,,
\end{equation}
where $R=R(T,x)$ and $\tilde{R} = \tilde{R}( \tilde{\mathcal{T}}, \tilde{\mathdutchcal{x}}) $ are arbitrary functions. We introduce the shorthand notation $R_{,\mathdutchcal{x}} := \frac{\partial R(\mathcal{T},\mathdutchcal{x})}{\partial \mathdutchcal{x}}$, with an analogous definition for $\tilde{R}_{,\mathdutchcal{\tilde{x}}}$. Considering the tangent vector $l^\alpha \partial_\alpha = \partial_T$, we find that it satisfies
\begin{equation}
l_{; \beta}^\alpha l^\beta=0\,, \quad l^\alpha l_\alpha=-1\,.
\end{equation}
Thus, $l^\alpha$ represents the tangent vector to a radial timelike geodesic parametrized by the proper time $T$. To determine the normal vector $n^\alpha$, we impose the normalization and orthogonality conditions
\begin{equation}
    n^\alpha n_\alpha=1\,, \quad n^\alpha l_\alpha=0\,,
\end{equation}
which yield the following components
\begin{equation} \label{normal}
    n^\alpha \partial_\alpha = \pm \frac{1}{R_{,x}} \partial_x \,.
\end{equation}
For the remainder of this analysis, we select the positive sign in Eq. \eqref{normal}\footnote{In applications to the OS collapse scenario, this ensures the normal vector points outward from the dust ball.}. We transform the tangent vector components into the coordinate system of \eqref{met2_glu} using the standard transformation law $l^{\tilde{\alpha}} = \frac{\partial y^{\tilde{\alpha}}}{\partial y^\alpha} l^\alpha$, where $y^\alpha$ and $y^{\tilde{\alpha}}$ denote the respective coordinate systems defined in \eqref{met1_glu} and \eqref{met2_glu}. The transformed components of the tangent vector are given by
\begin{equation}
    l^{\tilde{\alpha}} \partial_{\tilde{\alpha}} = \tilde{\mathcal{T}}_{, T} \partial_{\tilde{\mathcal{T}}} +  \tilde{\mathdutchcal{x}}_{, T} \partial_{\tilde{\mathdutchcal{x}}} \,,
\end{equation}
where the transformation functions $\tilde{\mathcal{T}}=\tilde{\mathcal{T}}(T, x)$ and $\tilde{\mathdutchcal{x}}=\tilde{\mathdutchcal{x}}(T, x)$ map the original coordinates to the new chart. Similarly, the components of the normal vector transform as
\begin{equation}
    n^{\tilde{\alpha}} \partial_{\tilde{\alpha}} =  \tilde{\mathcal{T}}_{, \mathdutchcal{x}} \frac{1}{R_{,x}}  \partial_{\tilde{\mathcal{T}}} +  \tilde{\mathdutchcal{x}}_{, \mathdutchcal{x}} \frac{1}{R_{,x}}  \partial_{\tilde{\mathdutchcal{x}}} \,.
\end{equation}
Enforcing geometric consistency via the normalization and orthogonality relations
\begin{equation}
l^{\tilde{\alpha}} l_{\tilde{\alpha}}=-1 \,, \quad n^{\tilde{\alpha}} n_{\tilde{\alpha}}=1 \,, \quad l^{\tilde{\alpha}} n_{\tilde{\alpha}}=0 \,,
\end{equation}
yields a system of three independent equations
\begin{equation} \label{jeq1}
    -\left(\tilde{\mathcal{T}}_{, T}\right)^2+\left(\tilde{\mathdutchcal{x}}_{, T}\right)^2\left( \tilde{R}_{, \tilde{\mathdutchcal{x}}} \right)^2=-1 \,,
\end{equation}
\begin{equation} \label{jeq2}
-\left(\tilde{\mathcal{T}}_{, x}\right)^2+\left(\tilde{\mathdutchcal{x}}_{, x}\right)^2\left(\tilde{R}_{, \tilde{\mathdutchcal{x}}}\right)^2=\left(R_{, x}\right)^2 \,,
\end{equation}
\begin{equation} \label{jeq3}
   -\tilde{\mathcal{T}}_{,T} \tilde{\mathcal{T}}_{,x}   + \tilde{\mathdutchcal{x}}_{,T}\tilde{\mathdutchcal{x}}_{,x}  \left(\tilde{R}_{, \tilde{\mathdutchcal{x}}}\right)^2=0 \,,
\end{equation}
Even if the metric functions $R_{,x}$ and $\tilde{R}_{,\tilde{\mathdutchcal{x}}}$ are known from the underlying dynamical equations, the resulting system of three equations, \eqref{jeq1}--\eqref{jeq3}, is generally underdetermined. Consequently, it cannot be solved algebraically for the four independent coordinate derivatives $\tilde{\mathcal{T}}_{, T}$, $\tilde{\mathcal{T}}_{, x}$, $\tilde{\mathdutchcal{x}}_{, T}$, and $\tilde{\mathdutchcal{x}}_{, x}$ without an additional constraint. For this reason, we calculate the induced metric on the $x=\mathrm{const.}$ hypersurface, which we denote as $\Sigma$ (representing the geodesic corresponding to $l^\alpha$). This yields
\begin{equation} \label{induced1}
    \dd \mathdutchcal{s}^2 \Big|_\Sigma =   -\dd T^2 +R^2 \dd \Omega^2 \,,
\end{equation}
and
\begin{equation} \label{induced2}
    \dd \tilde{\mathdutchcal{s}}^2\Big|_\Sigma = \left[ -\left(\tilde{\mathcal{T}}_{,T}\right)^2 + \left(\tilde{\mathdutchcal{x}}_{,T}\right)^2  \left(  \tilde{R}_{,\tilde{\mathdutchcal{x}}}  \right)^2\right]\dd T^2 +\tilde{R}^2 \dd \Omega^2 \,,
\end{equation}
where we have assumed $\dd \Omega^2 = \dd \tilde{\Omega}^2$, which is a natural identification under spherical symmetry. Comparing the $TT$-components of the induced metrics in \eqref{induced1} and \eqref{induced2} yields a condition identical to \eqref{jeq1}. Next, equating the angular components provides the condition
\begin{equation} \label{radii_junction}
    R(T,x)=\tilde{R}(\tilde{\mathcal{T}}, \tilde{\mathdutchcal{x}}) \,.
\end{equation}
To complete the system of equations \eqref{jeq1}--\eqref{jeq3}, we take the proper-time derivative of this relation, $\frac{\dd}{\dd T} R(T,x) = \frac{\dd}{\dd T} \tilde{R}(\tilde{\mathcal{T}}, \tilde{\mathdutchcal{x}})$, which expands via the chain rule into
\begin{equation} \label{radii_derivative}
    R_{,T} = \tilde{R}_{,\tilde{\mathcal{T}}} \tilde{\mathcal{T}}_{, T} +  \tilde{R}_{,\tilde{\mathdutchcal{x}}} \tilde{\mathdutchcal{x}}_{, T} \,.
\end{equation}
Enforcing this differentiated constraint ensures that the areal radius remains continuous across the junction hypersurface. The addition of relation \eqref{radii_derivative} closes the underdetermined system \eqref{jeq1}--\eqref{jeq3}. Indeed, we can now algebraically solve for the coordinate derivatives, yielding
\begin{equation} \label{dyn_junction1}
\tilde{\mathcal{T}}_{,T} =
\frac{
R_{,T}  \tilde{R}_{,\tilde{\mathcal{T}}}
\pm
\mathrm{sgn}\left(\tilde{R}_{,\tilde{\mathcal{T}}} \tilde{R}_{,\tilde{\mathdutchcal{x}}}\right)
\sqrt{
1
+
\left(R_{,T}\right)^2 -
\left(\tilde{R}_{,\tilde{\mathcal{T}}}\right)^2
}
}{
-1
+
\left(\tilde{R}_{,\tilde{\mathcal{T}}}\right)^2
} \,,
\end{equation}
\begin{equation} \label{dyn_junction2}
\tilde{\mathdutchcal{x}}_{,T}
= 
\frac{
- R_{,T}
\mp \tilde{R}_{,\tilde{T}} \,
\mathrm{sgn}\!\left(\tilde{R}_{,\tilde{\mathcal{T}}} \, \tilde{R}_{,\tilde{x}}\right)
\sqrt{
1
+
(R_{,T})^2
-
\left(\tilde{R}_{,\tilde{\mathcal{T}}}\right)^2
}
}{
R_{,x} \left[
(\tilde{R}_{,\tilde{\mathcal{T}}})^2
-1 \right] 
} \,,
\end{equation}
while the analogous expressions for $\tilde{\mathcal{T}}_{,x}$ and $\tilde{\mathdutchcal{x}}_{,x}$ are omitted here for brevity. In the subsequent analysis, we restrict our attention to the plus-sign branch in Eqs. \eqref{dyn_junction1} and \eqref{dyn_junction2}. The minus-sign variant typically yields a trivial junction where $\tilde{\mathcal{T}}_{,T} \to 1$ and $\tilde{\mathdutchcal{x}}_{,T} \to 0$ (after appropriately constraining the signs of the components on the right-hand sides of these expressions). This trivial case simply corresponds to matching with an adjacent geodesic, rather than describing a genuinely non-geodesic layer. Note that this trivial junction is viable at the surface of the dust ball before the bounce---see Fig. \ref{fig_shells}.

For the subsequent discussion, we also present the continuity condition for the components of the extrinsic curvature, defined by $K_{ab} = h_{\alpha \beta} \frac{\partial y^\alpha}{\partial z^a} \frac{\partial y^\beta}{\partial z^b}$. Here, $y^\alpha$ denotes the coordinates in either \eqref{met1_glu} or \eqref{met2_glu}, while $z^a = (T, \theta, \phi)$ represents the coordinates intrinsic to the hypersurface used for the induced metrics in \eqref{induced1} and \eqref{induced2}. Evaluating the continuity of the $\theta\theta$-component, namely $K_{\theta \theta} = \tilde{K}_{\theta \theta}$, yields the relation
\begin{equation} \label{extrinsic}
R_{,x}=\tilde{\mathdutchcal{x}}_{, x}  \tilde{R}_{,\tilde{\mathdutchcal{x}}}+\tilde{\mathcal{T}}_{, x} \tilde{R}_{,\tilde{\mathcal{T}}} \,.
\end{equation}
This relation will serve to test whether the extrinsic curvature remains continuous at the dust ball surface in the asymmetric OS collapse model.

\subsection{Detailed analysis of the dust ball surface} \label{secsub_detailed}

Having at our disposal junction equations derived in Sec. \ref{secsub_junction} we now apply them to investigate the properties of the surface of a dust ball in the asymmetric OS collapse scenario. Specifically, we establish the following identification: the external vacuum shells illustrated in Fig. \ref{fig_shells} are described by the line element \eqref{met2_glu}, while the interior LTB shells correspond to \eqref{met1_glu}. We assume that the junction hypersurface in the post-bounce spacetime region coincides with the surface of the collapsing dust ball. This choice is well-motivated since the vacuum shells carry no energy, so one should not expect extraordinary phenomena, that is, shocks, in the vacuum outside the dust ball. A direct consequence of this assumption is that the shock is located precisely at the boundary of the dust ball. Whether an alternative choice of junction surface can yield a self-consistent description compatible with the LTB dynamical equations remains an open question. 

Our objective is to determine the coordinate trajectories $\tilde{\mathcal{T}}$ and $\tilde{\mathdutchcal{x}}$ at the surface of the dust ball, parameterized by $T$. These trajectories will tell us, in particular, the exact times at which the vacuum shells impact the collapsing dust ball. At the junction surface $x=x_b$, this condition simplifies the coupled partial differential equations (PDE) given in \eqref{dyn_junction1} and \eqref{dyn_junction2} into a system of coupled ordinary differential equations (ODE) with $T$ as the time parameter. The functions $R_{,T}(T,x_b)$ and $R_{,x}(T,x_b)$ are known exactly at $x=x_b$ thanks to the solution in Eq. \eqref{Rsol}. In contrast, the transformed derivatives evaluated at the boundary, namely $\tilde{R}_{,\tilde{\mathcal{T}}} \big|_{x=x_b} = \tilde{R}_{,\tilde{\mathcal{T}}} [\tilde{\mathcal{T}}(T,x_b) , \tilde{\mathdutchcal{x}}(T,x_b)]$ and $\tilde{R}_{,\tilde{\mathdutchcal{x}}} \big|_{x=x_b} = \tilde{R}_{,\tilde{\mathdutchcal{x}}}[\tilde{\mathcal{T}}(T,x_b) , \tilde{\mathdutchcal{x}}(T,x_b)]$, are fully determined provided that the coordinate trajectories $\tilde{\mathcal{T}}(T,x_b)$ and $\tilde{\mathdutchcal{x}}(T,x_b)$ are known---the function $\tilde{R}(\tilde{\mathcal{T}}, \tilde{\mathdutchcal{x}})$ is again given by Eq. \eqref{Rsol}. We find these trajectories by solving the above-mentioned ODE system. 

To solve the system of equations \eqref{dyn_junction1} and \eqref{dyn_junction2} evaluated at $x=x_b$, appropriate initial conditions must be specified. Although the bounce point constitutes the most natural initial time $T=T_\mathrm{bounce}$, we encounter a numerical obstruction: the ratio $\tilde{R}_{,\tilde{\mathcal{T}}} / R_{,x}$ appearing in \eqref{dyn_junction2} cannot be evaluated there because $R_{,x} \to 0$ while $\tilde{R}_{,\tilde{\mathcal{T}}}$ is not yet known at the vicinity of that point. To circumvent this issue, we employ a standard blind-shooting method to numerically solve the ODE system, implemented as follows. First, we select a time $T=T^\mathrm{bs}_1$ shortly after the bounce at $x=x_b$, but in its immediate vicinity. With this time fixed, there are two initial values to determine—namely $\tilde{\mathcal{T}}(T^\mathrm{bs}_1)$ and $\tilde{\mathdutchcal{x}}(T^\mathrm{bs}_1)$—which serve as the targets for the shooting algorithm. Although this shooting parameter space is inherently two-dimensional, we can reduce it to a one-dimensional problem by enforcing the areal radius continuity condition \eqref{radii_junction}. Therefore, we use $\tilde{\mathdutchcal{x}}(T^\mathrm{bs}_1)$ as the sole shooting parameter, with $\tilde{\mathcal{T}}(T^\mathrm{bs}_1)$ fixed by the aforementioned continuity condition. The correct value for $\tilde{\mathdutchcal{x}}(T^\mathrm{bs}_1)$ is then identified by requiring the trajectory to satisfy $\tilde{\mathdutchcal{x}}(T_\mathrm{bounce}) = x_b$ at the bounce point. This procedure yields the unique coordinate trajectories $\tilde{\mathcal{T}}(T,x_b)$ and $\tilde{\mathdutchcal{x}}(T,x_b)$ that solve the ODE system \eqref{dyn_junction1} and \eqref{dyn_junction2}.

The numerical solution, generated using libraries of \texttt{Julia} programming language, is presented in Fig. \ref{fig_shells}. It displays the areal radius $\tilde{R}( \tilde{\mathcal{T}}(T,x_b), \tilde{\mathdutchcal{x}}(T,x_b))$ evaluated along the newly computed coordinate trajectories. As shown, the areal radius begins at the bounce point $R(T_\mathrm{bounce},x_b)$ and rapidly asymptotes toward the constant line $R=R_2$ as $T \to \infty$. Each intersection between a vacuum shell and this trajectory marks the specific time $\tilde{\mathcal{T}}$—readable via the corresponding $T$-coordinate on the horizontal axis—at which the shell impacts the surface of the dust ball. Evidently, every vacuum shell impacts the surface at a radius smaller than $R=R_2$. This confirms a nontrivial property already present in the symmetric bounce scenario: at the instant of impact, a discontinuity arises between the proper time measured by an observer comoving with the dust ball's surface and that of successive observers comoving with the vacuum shells. While this property may initially seem counterintuitive, it becomes much more transparent when viewed on the conformal diagram of the collapse, which we present in Sec. \ref{sec:causal}.

\begin{figure}
    \centering
    \includegraphics[width=1\linewidth]{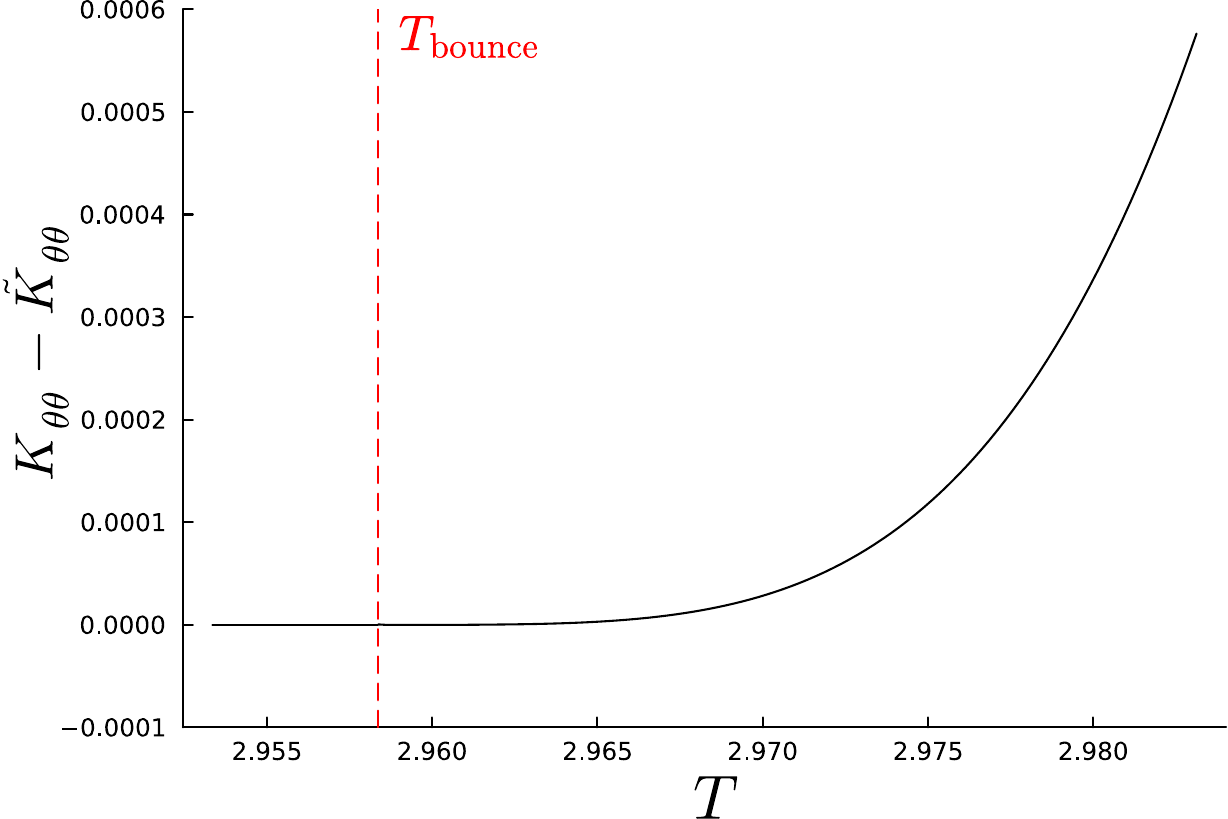}
    \caption{The discontinuity of the extrinsic curvature, $K_\mathrm{\theta\theta} - \tilde{K}_\mathrm{\theta\theta}$, at the surface of the collapsing dust ball, parameterized by $T$, according to Eq. \eqref{extrinsic}. Prior to the bounce ($T < T_\mathrm{bounce}$), we have $\tilde{\mathdutchcal{x}}_{,x} \to 1$ and $\tilde{\mathcal{T}}_{,x} \to 0$. In this regime, Eq. \eqref{extrinsic} trivializes, yielding a continuous extrinsic curvature component. However, a discontinuity emerges immediately after the bounce ($T > T_\mathrm{bounce}$). The parameters used here are identical to those in Fig. \ref{fig_shells} (in Planck units).  }
    \label{fig:discontinuity}
\end{figure}

Once the functions $\tilde{\mathcal{T}}(T,x_b)$ and $\tilde{\mathdutchcal{x}}(T,x_b)$ are determined at the junction surface, we can evaluate the continuity condition for the extrinsic curvature component given in Eq. \eqref{extrinsic}. Our numerical calculations---Fig. \ref{fig:discontinuity}---confirm that this component is discontinuous after the bounce for the solutions presented in Fig. \ref{fig_shells}. This is not an unexpected result, however, given the underlying dynamics. As shown in Fig. \ref{fig_shells}, every vacuum shell impacts the dust ball's surface before undergoing its own bounce. This means that at the moment of impact, the vacuum shells are still in a power-law dynamical phase, whereas the dust ball surface has already transitioned into a de Sitter-like expanding phase. This kinematic mismatch between the two regions causes the discontinuity in the extrinsic curvature. Such a discrepancy is absent in the symmetric OS collapse scenario, where the power-law contraction of the dust ball surface is seamlessly followed by a power-law expansion. Ultimately, this discontinuity in the extrinsic curvature proves that a weak $C^0$ shock forms at the surface of the dust ball during the post-bounce regime. 

\section{Causal structure analysis} 
\label{sec:causal}

Having investigated the nature of the junction (dust ball) surface we can now proceed with determining the exact global causal structure of the investigated collapse scenario. To this end, we employ the techniques developed in our previous works \cite{Bobula:2024ywp,Bobula:2024chr} to numerically compute the exact Penrose-Carter (conformal) diagram. The resulting causal structure of the asymmetric Oppenheimer--Snyder collapse is shown in Fig.~\ref{fig:asymmetric_collapse}. 

The construction proceeds in several successive steps. First, Sec. \ref{subsec_affine} establishes the affine parametrization for vacuum radial null geodesics corresponding to LTB-like solution in Eq. \eqref{Rsol} and identifies the areal radii of the trapping horizons. Next, Sec. \ref{sec_sub_null_propagated} uses these results to probe the causal structure via null geodesics propagated from the surface of the dust ball. Following this, Sec. \ref{sec_sub_beyond_min} analyzes the non-trivial continuation of those null geodesics that reach the minimal-radius surface $R_\mathrm{min}$. We then determine the trapping horizons within the homogeneous and isotropic interior of the dust ball in Sec. \ref{subsec_interior}. Finally, Sec. \ref{subsec_conformal} constructs the global double-null coordinate chart covering the complete asymmetric OS collapse and discusses the resulting conformal diagram.

\subsection{Affine parametrization of null geodesics and horizon radii} \label{subsec_affine}

We begin by analyzing null geodesics in the exterior vacuum region, where $M(x)=M=\mathrm{const.}$. Specifically, we first determine their affine parametrization and then identify the values of the areal radius corresponding to the horizons of the exterior vacuum spacetime. This analysis provides the foundation for accurately locating these horizons and probing the spacetime extensions.

As the solution to the quantum-corrected LTB equations of motion is expressed in an $\eta$-parametric form in Eq. \eqref{Rsol}, it is advantageous to transform the time coordinate as $T \to \eta$ in the line element \eqref{met}. This coordinate change provides a more suitable basis for investigating potential analytic extensions of the collapse spacetime. Taking the differential of the second equation in \eqref{Rsol} yields
\begin{equation}
    s'(x)\, \mathrm{d}x - \mathrm{d}T 
+ \frac{4\alpha_\Delta^2\gamma^2 + 9\eta^2}
{4\alpha_\Delta^2\gamma^4 - 9\eta^2}\, \mathrm{d}\eta = 0 \,,
\end{equation}
Solving this for $\mathrm{d}T$ allows us to transform the line element \eqref{met} into 
\begin{equation} \label{met_eta_x}
\begin{aligned}
\mathrm{d}s^2 = -\frac{(4\alpha_\Delta^2\gamma^2 + 9\eta^2)^2}{(4\alpha_\Delta^2\gamma^4 - 9\eta^2)^2} \mathrm{d}\eta^2 - \frac{2(4\alpha_{\Delta}^2\gamma^2 + 9\eta^2)s'(x)}{4\alpha_{\Delta}^2\gamma^4 - 9\eta^2} \mathrm{d}x \mathrm{d}\eta \\ +    \left[ \left( \frac{3 \, G M \eta \left( 4\alpha_{\Delta}^2(\gamma^2 + 2\gamma^4) - 9\eta^2 \right)}{R^2 \left(4\alpha_{\Delta}^2\gamma^4 - 9\eta^2\right)} \right)^2 - 1 \right]  s'(x)^2 \dd x^2 \\ +  R^2 \mathrm{d}\Omega^2 \,,
\end{aligned}
\end{equation}
where $R$ is given by Eq. \eqref{Rsol} with a constant $M$. Although it is tempting to further transform $x \to R$ in Eq. \eqref{met_eta_x} because the areal radius $R$ carries a clear geometric significance, the Jacobian of this transformation is singular. Imposing the null condition $\mathrm{d}s^2 = 0$, we find
\begin{equation} \label{null_conds}
\begin{aligned}
  &\left( \frac{\mathrm{d}x}{\mathrm{d}\eta} \right)_\pm= \\  &\frac{ \frac{2 R^3}{GM}  \left(4\alpha_\Delta^2\gamma^2 + 9\eta^2\right)}{s'(x) \left[ \left(4\alpha_\Delta^2\gamma^2 + 9\eta^2\right)^2 \pm 6 R \eta \left(4\alpha_\Delta^2\gamma^2(1+2\gamma^2) - 9\eta^2\right) \right]} \,.
\end{aligned}
\end{equation}
This allows us to construct two radial null vectors, ${}^{\pm}l^\alpha$, satisfying ${}^{\pm} l^\alpha {}^{\pm} l_\alpha =0$, which are expressed as
\begin{equation}
    {}^\pm l^\alpha \partial_\alpha = \pm \partial_\eta \pm \left( \frac{\mathrm{d}x}{\mathrm{d}\eta} \right)_\mp \partial_x \,.
\end{equation}
The null vectors components ${}^{+} l^\alpha$ and ${}^{-} l^\alpha$ are parametrized by $\eta$ and $-\eta$ respectively. The vectors satisfy
\begin{equation}
    {}^\pm l^\alpha{}_{; \beta} {}^\pm l^\beta = \kappa_\pm  {}^\pm l^\alpha{} \,,
\end{equation}
with 
\begin{equation} \label{kappas}
\begin{aligned}
\kappa_\pm &= \pm \frac{1}{\eta} + 6\eta \Biggl( \mp \frac{1}{9\eta^2+4\alpha_\Delta^2\gamma^2} \\
&\quad \mp \frac{4}{9\eta^2-4\alpha_\Delta^2\gamma^4} \pm \frac{3}{9\eta^2-4\alpha_\Delta^2\gamma^2(1+2\gamma^2)} \Biggr) \,.
\end{aligned}
\end{equation}
Consequently, the studied null vectors generate radial null geodesics. To find affine parameters on these geodesics $\lambda_\pm$ we solve the differential equations (see e.g. Ch. 1.3 of  \cite{Poisson:2009pwt} for a general discussion)
\begin{equation} \label{affine_eqs}
    \frac{\mathrm{d}^2 \lambda_\pm }{ \mathrm{d}\eta^2 } = \kappa_\pm  \frac{\mathrm{d} \lambda_\pm }{ \mathrm{d}\eta } \,.
\end{equation}
We find the solutions 
\begin{equation} \label{affinesol}
\lambda_\pm(\eta) =c_2
+
c_1\, R \,,
\end{equation}
where $c_1$ and $c_2$ are the integration constants. The form of the solution \eqref{affinesol} shows that areal radius $R$ is an affine parameter on the null geodesics in the vacuum sector $M=\mathrm{const.}$ of the studied geometry. This is expected result for the spherically symmetric vacuum.

\begin{figure}
    \centering
    \includegraphics[width=\linewidth]{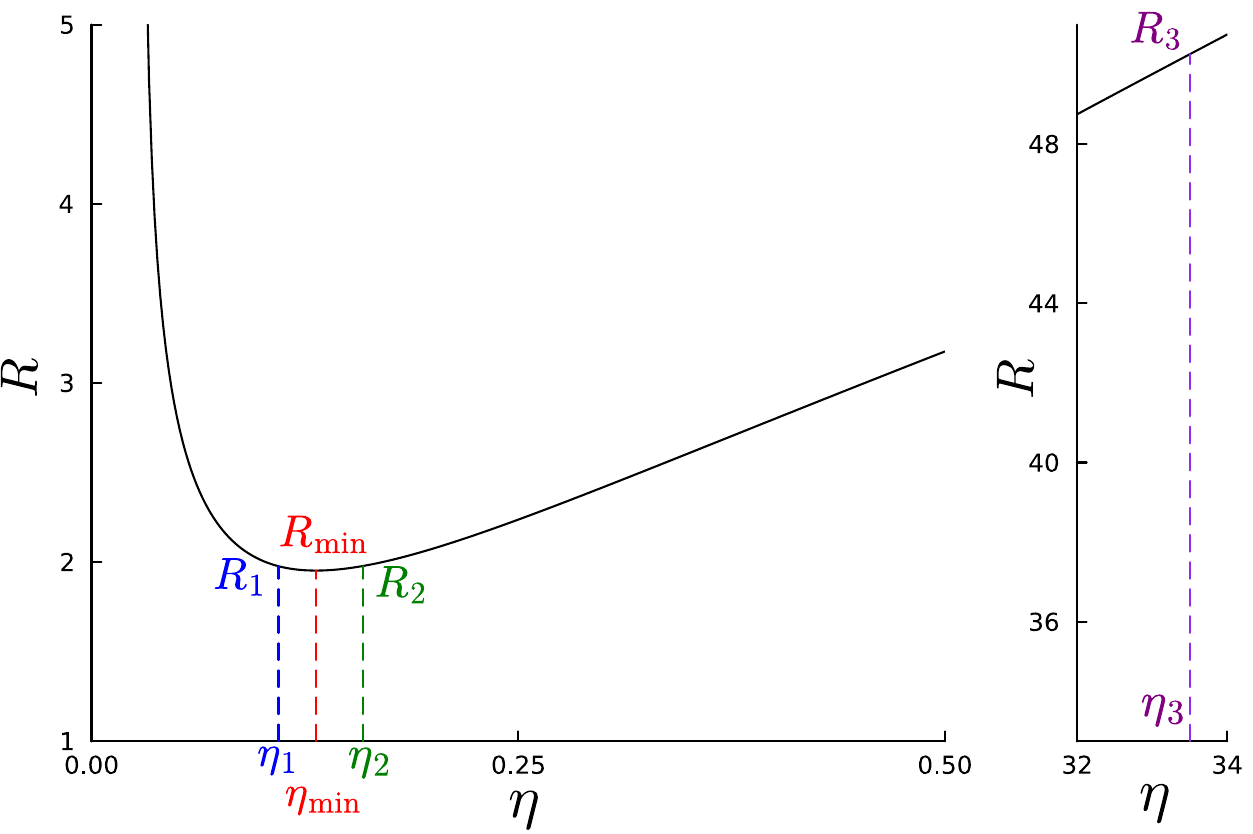}
    \caption{The areal radius $R(\eta)$ valid for the vacuum sector of the solution given in Eq. \eqref{Rsol}. We have $R(\eta_\mathrm{inf}) \to \infty$ on the left branch of the curve, while $R(\eta \to \infty) \to \infty$ on the right branch. The minimal radius $R_\mathrm{min}$ and its corresponding $\eta_\mathrm{min}$ are marked. Additionally, the positions of the null horizons $R_1$, $R_2$, and $R_3$ are indicated (compare with the conformal diagrams in Figs. \ref{fig:asymmetric_collapse} and \ref{fig:conformal_vacuum}). This plot was generated using the same parameters (in Planck units) as in Fig. \ref{fig_shells}. }
    \label{fig:Reta}
\end{figure}

To find areal radii at the trapping horizons, we exploit certain universal formula found for spherically symmetric spacetimes to be mathematically equivalent to a condition, that a surface is foliated by marginally trapped slices \cite{Hayward:2000ca, Faraoni:2016xgy}
\begin{equation} \label{appa_cond}
R_{; \alpha}R^{;\alpha} = 0\,,
\end{equation}
which we evaluate using the line element \eqref{met_eta_x}. We restrict our analysis to collapsing dust ball of sufficiently high mass, which naturally confines the system to non-extremal black hole configurations. We find three distinct positive roots of Eq. \eqref{appa_cond}, ordered such that $\eta_1 < \eta_\mathrm{min} < \eta_2 < \eta_3$. These $\eta_i$ correspond to the roots of the equation $\left\vert{} 9\eta^2 - 4\alpha^2\gamma^4 \right\vert{} (9\eta^2 + 4\alpha^2\gamma^2)^4 = 108 G M \eta^3 \left\vert{} 9\eta^2 - 4\alpha^2\gamma^2(1+2\gamma^2) \right\vert{}^3$. Substituting these roots into Eq. \eqref{Rsol} yields the corresponding values for the areal radius, which obey the hierarchy $R_{\mathrm{min}} < R_1 < R_2 < R_3$. Eq. \eqref{appa_cond} possesses an additional root, namely $\eta_\mathrm{inf} = \frac{2}{3} \alpha_\Delta \gamma^2 < \eta_1$, however, this corresponds to the limit $R \to \infty$. Because this value of the areal radius is reached only at infinite affine time along both timelike (see Fig. \ref{fig_shells}) and null geodesics (see Eq. \eqref{affinesol}), it does not describe a null horizon. Rather, it corresponds to spacelike infinities, $R\to \infty$, as depicted in Fig. \ref{fig:asymmetric_collapse}. In Fig. \ref{fig:Reta}, we plot the areal radius $R$ as a function of $\eta$, as given by Eq. \eqref{Rsol}, valid in the exterior vacuum region. Indeed, the surfaces of constant areal radius $R = R_1$, $R_2$, and $R_3$ correspond to null trapping horizons.
\subsection{Null geodesics propagated from the surface of the dust ball} \label{sec_sub_null_propagated}
A key step in analyzing the causal structure is to determine the behavior of affine parameters along null geodesics. Specifically, we examine null geodesics emerging outward from the surface of the collapsing dust ball prior to the bounce (see the conformal diagram in Fig. \ref{fig:asymmetric_collapse}). This step inform us where exactly the trapping horizons are located in the exterior vacuum. 

We evaluate the null conditions \eqref{null_conds} at the surface of the dust ball during the pre-bounce phase ($\eta > \eta_\mathrm{min}$). We find that $\left( \frac{\dd x}{\dd \eta } \right)_+ \Big\vert{}_{x=x_b} > 0$ for all $\eta > \eta_\mathrm{min}$, equivalently $R > R_\mathrm{min}$. In contrast, $\left( \frac{\dd x}{\dd \eta } \right)_-$ behaves as follows
\begin{enumerate}[label=(\roman*)]
\item $\left( \frac{\dd x}{\dd \eta } \right)_- \Big\vert{}_{x=x_b} > 0$ for $\eta \in (\eta_3, \infty)$.
\item $\left( \frac{\dd x}{\dd \eta } \right)_- \Big\vert{}_{x=x_b} < 0$ for $\eta \in (\eta_2, \eta_3)$.
\item $\left( \frac{\dd x}{\dd \eta } \right)_- \Big\vert{}_{x=x_b} > 0$ for $\eta \in (\eta_\mathrm{min}, \eta_2)$.
\end{enumerate}
For any future-directed light ray, the coordinate $x$ must decrease as the ray penetrates the interior of the dust ball, whereas $x$ increases for outward-moving rays. Combining these kinematic properties with the plot of $R(\eta)$ in Fig. \ref{fig:Reta}, we can conclude that $\left( \frac{\dd x}{\dd \eta } \right)_+$ describes future-directed null geodesics entering the dust ball, while $\left( \frac{\dd x}{\dd \eta } \right)_-$ describes those moving outward. Based on these observations, we further conclude that for future-directed null geodesics moving outward from the surface of the dust ball at an initial radius:
\begin{enumerate}[label=(\roman*)]
\item $R\in (R_3,\infty)$, $R$ increases to infinity and no horizon is crossed.
\item $R\in (R_2,R_3)$, $R$ decreases to $R_\mathrm{min}$ while the $R_2$ horizon is crossed.
\item $R\in (R_\mathrm{min},R_2)$, $R$ increases to infinity while the $R_2$ and $R_3$ horizons are crossed.
\end{enumerate}
Indeed, once the monotonicity of the areal radius (the affine parameter) is specified at the initial point, it must remain the same along a given null geodesic. This holds at least up to the value $R_\mathrm{min}$, as occurs in the second case above. Extending the geodesics beyond this value requires the additional discussion provided in Sec. \ref{sec_sub_beyond_min}. The observations presented here can be compared with the conformal diagram shown in Fig. \ref{fig:asymmetric_collapse}.

\subsection{Spacetime extension beyond minimal radius $R_\mathrm{min}$} 
\label{sec_sub_beyond_min}

As shown in Sec. \ref{sec_sub_null_propagated}, there exist null geodesics traced toward the future from the dust ball surface that approach $R_\mathrm{min}$ (see also the conformal diagram in Fig. \ref{fig:asymmetric_collapse}). It is then both natural and nontrivial to ask, whether they can be extended beyond this point. We will address this question now.

At first glance, one might expect the areal radius $R$ to continue decreasing beyond $R_\mathrm{min}$, since we stated earlier that its monotonicity should be preserved. As an aside, this is exactly the case for the symmetric OS collapse scenario \cite{Husain:2021ojz, Husain:2022gwp, Lewandowski:2022zce, Bobula:2023kbo, Fazzini:2023scu, Bobula:2024chr}, where the areal radius decreases up to the timelike singularity located at areal radius equal to zero. Here for the asymmetric collapse, the situation is however very different. 

First, we observe that at $\eta_\mathrm{min}$ (or equivalently $R_\mathrm{min}$) the quantities \eqref{kappas} diverge. Since they appear in the differential equation \eqref{affine_eqs}, the standard existence and uniqueness conditions for ODEs are violated at this point. Therefore, the solution \eqref{affinesol} does not admit a unique continuation beyond $R_\mathrm{min}$. Thus, when null geodesics are propagated, e.g. from the dust-ball surface towards the future and decreasing $R$, the areal radius provides a valid affine parametrization only in the range $R > R_\mathrm{min}$.

To find the appropriate extension for the discussed geodesics, we rewrite the null condition $\left( \frac{\dd x}{\dd \eta } \right)_-$ in the integral form 
\begin{equation} \label{null_int}
\begin{aligned}
     &s(x) - s(x_\mathrm{in}) =  \\ &\int_{\eta_\mathrm{in}}^\eta \frac{ \frac{2 R^3}{GM}  \left(4\alpha_\Delta^2\gamma^2 + 9\eta^2\right)}{ \left[ \left(4\alpha_\Delta^2\gamma^2 + 9\eta^2\right)^2 - 6 R \eta \left(4\alpha_\Delta^2\gamma^2(1+2\gamma^2) - 9\eta^2\right) \right]} \,, 
\end{aligned}
\end{equation}
where we introduced the initial condition $x(\eta_\mathrm{in}) = x_\mathrm{in}$. In expression \eqref{null_int}, we utilize the vacuum variant of the function $s(x)$ provided in App. \ref{appendix_sx}, as we are seeking extensions of a vacuum spacetime. Specifically, this involves incorporating the formula from Eq. \eqref{sx_formula} alongside the second variant of Eq. \eqref{eta0} for all possible values of $x$. Consequently, the range of this coordinate is $x \in (x_s, \infty)$, where $x_s := 2 \sqrt[3]{\alpha_{\Delta}^2 \gamma^2\left(1+\gamma^2\right) M}$ is the smallest value of $x$ that ensures Eq. \eqref{sx_formula} remains real-valued.

\begin{figure}
    \centering
    \includegraphics[width=1\linewidth]{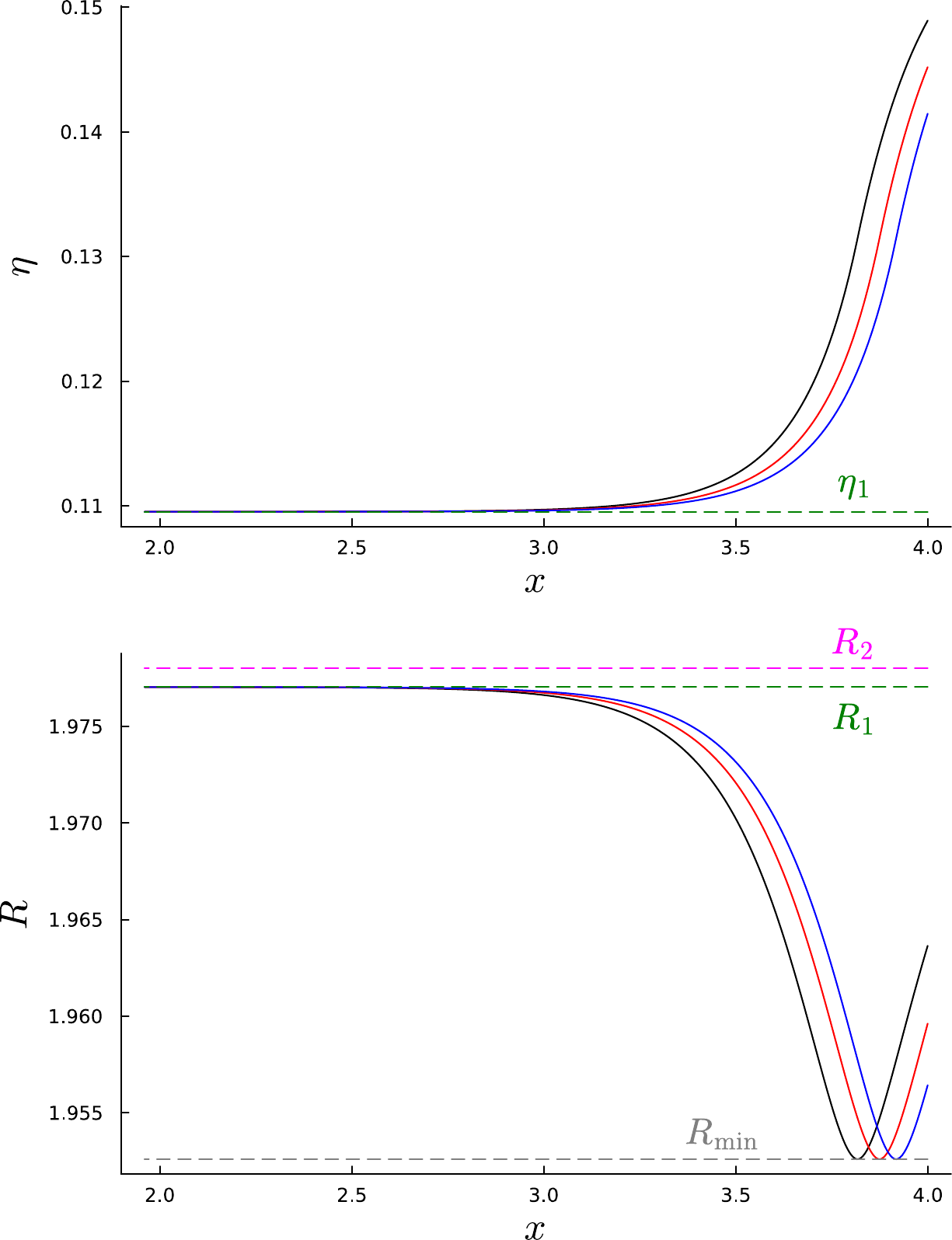}
    \caption{\emph{Top.} Behavior of $\eta$ versus $x$ along three distinct null geodesics approaching the minimal radius $R_\mathrm{min}$. The three functions $\eta(x)$, obtained as solutions to Eq. \eqref{null_int}, are plotted for a common $x_\mathrm{ini} = 4$ and different initial $\eta_\mathrm{ini} \in \{ \eta_2 -0.01, (\eta_2 + \eta_\mathrm{min})/2, \eta_\mathrm{min} +0.01 \}$, corresponding to the black, red, and blue curves, respectively. The coordinate $x$ ranges from $x_s$ to $x_\mathrm{ini}$ (where $x_\mathrm{ini} > x_s$). In the limit $x \to x_s$, all curves asymptotically approach $\eta_1$. The $\eta = \eta_1$ constant line is also plotted. Parameters are identical to those in Fig. \ref{fig_shells}, with $x_\mathrm{ini}=4$. Planck units are used throughout. \emph{Bottom.} The areal radius $R[\eta(x)]$ evaluated along the same three trajectories, where $\eta(x)$ corresponds to the solutions shown in the top panel (using matching colors) and $R(\eta)$ is defined by Eq. \eqref{Rsol} for the vacuum sector. When the geodesics are traced forward in time—along decreasing $x$—the areal radius first decreases, reaches a minimum at $R_{\mathrm{min}}$ (indicated by the gray dashed line), and subsequently increases, asymptotically approaching $R_1$ as $x \to x_s$. }
    \label{fig:geodesics}
\end{figure}

We now provide a detailed analysis of the future-directed, outward-propagating null geodesics emanating from the dust ball surface at $x=x_b$ for $\eta \in (\eta_2,\eta_3)$. According to Eq. \eqref{null_int}, $x$ increases while $\eta$ decreases along these trajectories. As $x \to \infty$ along each geodesic, the radius $R_2$ is reached. This limit corresponds to both the outermost vacuum shell and a Cauchy horizon of the spacetime (see the diagram in Fig. \ref{fig:asymmetric_collapse}). Because $x$ diverges to infinity, a new coordinate patch is required to describe the spacetime beyond this limit. According to Eq. \eqref{null_int}, the coordinate $x$ decreases from infinity in the region beyond the Cauchy horizon, while $\eta$ continues to decrease toward $\eta_\mathrm{min}$ (corresponding to $R_\mathrm{min}$). A detailed analysis of the coordinate behavior near this minimum radius is provided in Fig. \ref{fig:geodesics}. There, we plot the solutions $\eta(x)$ to Eq. \eqref{null_int} for a fixed $x_\mathrm{ini}$ and various initial values of $\eta_\mathrm{ini}$. These starting points are chosen to lie already beyond the Cauchy horizon. As shown in Fig. \ref{fig:geodesics}, along these null geodesics, the areal radius decreases toward $R_\mathrm{min}$ and subsequently increases after passing through it. We thus obtain an unambiguous extension of the null geodesics beyond $R_\mathrm{min}$, under the choice that $x$ remains monotonic across this minimum\footnote{Note that before crossing $R_{\rm min}$ the coordinate $x$ enumerates free-falling timelike vacuum shells (possibly intersecting some spacelike slice, on which $x$ has the meaning as in the Eq.~\eqref{R0}) which intersect $R=R_\mathrm{min}$ surface for finite $T$. This assigns values of $x$ to each point of that surface, thus carries the definition of $x$ to the region beyond $R=R_\mathrm{min}$, furthermore providing the $x$-parametrization of $R_\mathrm{min}$ surface with definite time orientation. }
This assumption is motivated by the fact that the $x$ coordinate labels adjacent timelike vacuum shells. As a consequence, the $\eta$ coordinate remain monotonic on these geodesics. Furthermore, the fact that the discussed geodesics reach the same value $R=R_1$ reflects that they reach the null horizon—see Fig. \ref{fig:asymmetric_collapse}—in the common limit $x \to x_s$.

The extension of null geodesics beyond $R_\mathrm{min}$ shown in Fig. \ref{fig:geodesics} has direct consequences for their affine parameterization. Prior to reaching $R_\mathrm{min}$, the areal radius $R$ serves as the affine parameter (up to multiplicative and additive constants, according to the minus sign variant of Eq. \eqref{affinesol}). However, after crossing $R_\mathrm{min}$, $-R$ becomes the affine parameter, with new $c_1$ and $c_2$ suitably chosen to ensure continuity with the solution prior to $R_\mathrm{min}$. 

We conclude this part with another observation—since $-R$ has to remain monotonic in order to preserve the monotonicity of the affine parameter, the areal radius along the null geodesics discussed here continues to increase toward $R\to \infty$ ($\eta \to \eta_\mathrm{inf}$). However, it encounters no further horizons, in accordance with the analysis in Fig. \ref{fig:Reta}. In this region beyond $x_s$, yet another $x$ coordinate patch is required. The analysis presented above imply that the $\lambda_-$ parameter remains globally monotonic.

\subsection{Interior of the dust ball} 
\label{subsec_interior}

Let us now turn our attention to the metric of the cosmological spacetime of the interior of the dust ball and locate its trapping horizons.

As the interior is spatially flat, homogeneous, and isotropic, the LTB-like metric (line element \eqref{met}) simplifies to
\begin{equation} \label{frw}
\mathrm{d} \mathdutchcal{s}^2=-\mathrm{d} T^2+a(T)^2 \mathrm{d} x^2+a(T)^2 x^2 \mathrm{d} \Omega^2 \,,
\end{equation}
where the coefficient $a(T)$ (the scale factor) can be expressed as
\begin{equation}
    a(T):= \frac{\partial R(T,x)}{\partial x} = R(T,x)/x \, .
\end{equation}
While an analytical verification of the right equality above is cumbersome, it was straightforward to verify numerically and was confirmed to within the numerical error. 

For the line element \eqref{frw}, the condition \eqref{appa_cond} for the location of trapping horizons in presence of a spherical symmetry yields
\begin{equation}
1-\dot{a}^2 x^2=0 \,.
\end{equation}
Fig. \ref{fig:asymmetric_collapse} illustrates the two trapping horizons present within the interior. The first trapping horizon enters the interior at $R_3$, approaches the vicinity of the center of the dust ball, and then returns to the surface, exiting at the areal radius $R_2$. The second trapping horizon enters the dust ball at $R_1$ and gradually approaches the center. Numerical analysis indicates that it reaches the center as $T \to \infty$.

\subsection{Conformal diagram} \label{subsec_conformal}
\begin{figure*}
    \centering
\includegraphics[width=1\linewidth]{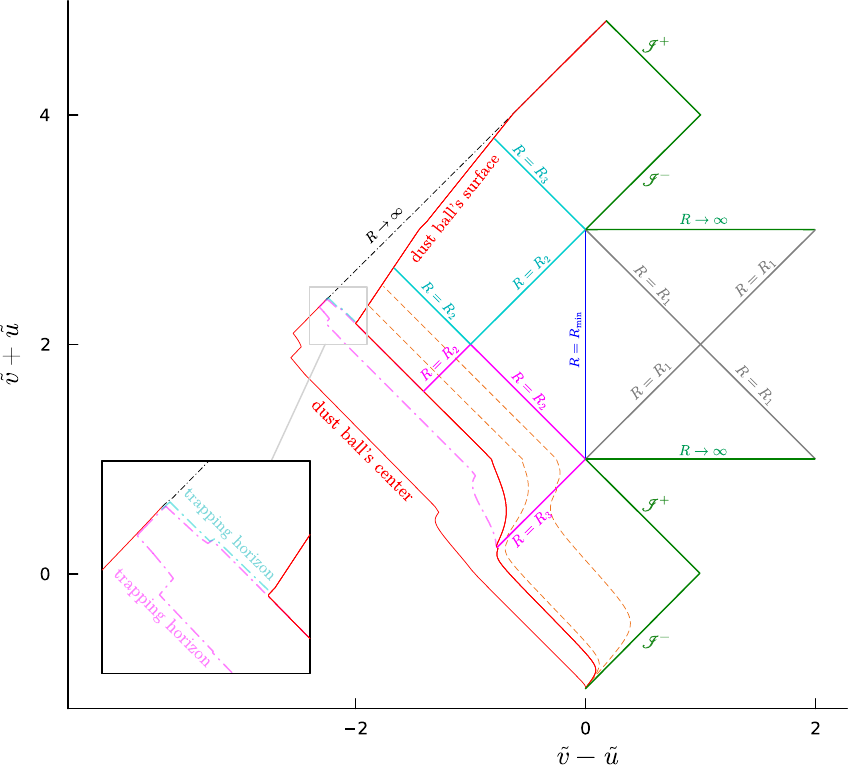}
    \caption{Numerically computed conformal diagram for the asymmetric OS collapse scenario. Prior to the bounce, the surface of the dust ball crosses a pair of null horizons, first $R_3$ and then $R_2$. These horizons are connected by a trapping horizon traversing the interior (pink dash-dotted curve). After the bounce, the dust ball surface becomes the physical shock surface. It passes a second trapping horizon in the interior (cyan dash-dotted curve) at a spacetime point distinct from where it crosses the subsequent $R_2$ null horizon in the post-bounce vacuum region. The surface then crosses the $R_3$ horizon and terminates at $\mathcal{J}^+$. The two gold dashed lines correspond to the vacuum shells from Fig. \ref{fig_shells}. They intersect the dust ball surface at a radius $R < R_2$. The interior spacelike infinity is represented by a black dash-dotted curve. Future and past null infinities are indicated by diagonal green lines. The spacetime admits a Cauchy horizon, which consists of the pink $R_2$ horizon evolving into the cyan $R_2$ horizon, whose extension enters the surface of the dust ball. Finally, the spacetime extension beyond the minimal radius $R_\mathrm{min}$ is shown, featuring a null horizons at radius $R_1$ alongside spacelike infinities ($R \to \infty$). To generate this diagram, we used the same parameters as in Fig. \ref{fig_shells}, along with $\kappa=0.1$, $T^\mathrm{bs}_2=3.108$, $\tilde{\mathdutchcal{x}}(T^\mathrm{bs}_2)=20$, $T^\mathrm{bs}_3=5.8230$, and $\tilde{\mathdutchcal{x}}(T^\mathrm{bs}_3)=15$ (in Planck units).   }
    \label{fig:asymmetric_collapse}
\end{figure*}

We now elaborate on how to construct double-null coordinates and compactify them to obtain the result presented in Fig. \ref{fig:asymmetric_collapse}, which depicts the conformal diagram for the asymmetric OS collapse scenario.

We begin by establishing double-null coordinates for the exterior vacuum. Specifically, we seek a viable double null pair $(u,v)$. Their corresponding differentials must satisfy $\dd (\dd u)=0$ and $\dd (\dd v) =0$, as this is a necessary condition for the existence of such a pair \cite{Bobula:2024chr}. To this end, we find that transitioning from the LTB-like coordinates in Eq. \eqref{met} to Painlev\'e-Gullstrand-like coordinates before writing $\dd u$ and $\dd v$ will be advantageous for our goal, since the discussed criteria are then easily met. We express the differential of the areal radius as
\begin{equation} \label{diffeR}
    \mathrm{d}R = \frac{\partial R(T,x)}{\partial T} \mathrm{d}T + \frac{\partial R(T,x)}{\partial x} \mathrm{d}x \,.
\end{equation}
where the components evaluated in the vacuum sector are given by
\begin{equation}
 \frac{\partial R(T,x)}{\partial T}=  \frac{6 R \eta \left( 4\alpha^2(\gamma^2 + 2\gamma^4) - 9\eta^2 \right)}{(4\alpha^2\gamma^2 + 9\eta^2)^2} \,,
\end{equation}
\begin{equation}
\frac{\partial R(T,x)}{\partial x}=    \frac{6 R\eta \left( 9\eta^2 - 4\alpha^2(\gamma^2 + 2\gamma^4) \right) s'(x)}{(4\alpha^2\gamma^2 + 9\eta^2)^2} \,.
\end{equation}
To substitute $\eta$, we invert the first expression in Eq. \eqref{Rsol} to obtain
\begin{equation} \label{etaR}
\eta(R)
=
\frac{1}{3} \sqrt{-4\alpha_\Delta^2 \gamma^2 + \frac{R^3}{GM} + \frac{\Xi(R)\sigma}{GM}}\,, \quad
\end{equation}
where $\sigma := \mathrm{sgn}\!\left(\frac{\mathrm{d}R}{\mathrm{d}\eta}\right)$ determines which branch of $R(\eta)$ is used for the inversion (see Fig. \ref{fig:Reta}). That is, the choice of $\sigma = \pm 1$ determines the vacuum regime, corresponding to either power-law contracting or exponentially expanding vacuum shells. In Eq. \eqref{etaR} we introduced 
\begin{equation}
\Xi(R):=\sqrt{R^3 \left(R^3 - 8\alpha_\Delta^2 G \gamma^2 M (1 + \gamma^2)\right)} \,.
\end{equation}
We are now ready to solve Eq. \eqref{diffeR} for $\mathrm{d}x$ and eliminate this differential from the line element \eqref{met}. This yields the metric in Painlev\'e--Gullstrand-like form
\begin{equation} \label{PGmet}
 \mathrm{d}s^2=   -\mathrm{d}T^2
+ 
\left(
\mathrm{d}R
+
N
\,\mathrm{d}T
\right)^2
+
R^2\,\mathrm{d}\Omega^2 \,,
\end{equation}
where the shift function is
\begin{widetext}
\begin{equation} \label{shift}
N 
:=
\frac{
2^{2/3}
\sqrt{
GM\left(
-4\alpha_\Delta^2GM\gamma^2
+R^3
+\Xi(R)\sigma
\right)
}
\left(
-8\alpha_\Delta^2GM\gamma^2(1+\gamma^2)
+R^3
+\Xi(R)\sigma
\right)
}{
\left(R^3+\Xi(R)\sigma\right)^{4/3}
\left(
-4\alpha_\Delta^2GM\gamma^2(1+\gamma^2)
+R^3
+\Xi(R)\sigma
\right)^{1/3}
} \,.
\end{equation}
\end{widetext}
The double-null coordinates can then be written as
\begin{equation}
\begin{aligned}
& u=T-\int_{R_\mathrm{min}}^{R} \frac{1}{1-N} \mathrm{~d} \tilde{R}\,, \\
& v=T+\int_{R_\mathrm{min}}^{R} \frac{1}{1+N} \mathrm{~d} \tilde{R}\,.
\end{aligned}
\end{equation}
These coordinates satisfy the existence criteria discussed earlier. We now turn our attention to the trajectory of the dust ball surface when expressed in these coordinates
\begin{equation} \label{null_traj}
\begin{aligned}
& u(T)=\tilde{\mathcal{T}}(T,x_b)-\int_{R_\mathrm{min}}^{R(T,x_b)} \frac{1}{1-N\big|_{\sigma=1}} \mathrm{~d} \tilde{R}\,, \\
& v(T)=\tilde{\mathcal{T}}(T,x_b)+\int_{R_\mathrm{min}}^{R(T,x_b)} \frac{1}{1+N\big|_{\sigma=1}} \mathrm{~d} \tilde{R}\,.
\end{aligned}
\end{equation}
Regarding these expressions, a few remarks are in order. Before the surface of the collapsing dust ball bounces, we have the simple case $\tilde{\mathcal{T}}(T,x_b) \to T$, where the proper time is continuous across the surface, as discussed in Sec. \ref{sec:analysis_of_asymmetric}. However, after the bounce, $\tilde{\mathcal{T}}(T,x_b)$ is determined by the solution to the ODE system \eqref{dyn_junction1}, \eqref{dyn_junction2} introduced earlier. Immediately following the bounce, the vacuum shells impacting the dust ball's surface remain in the power-law contracting dynamical regime. For this reason, we set $\sigma = 1$ in Eqs. \eqref{null_traj}. As $\tilde{\mathcal{T}} \to \infty$ at a finite time $T=:T_\mathrm{Cauchy}$ determined by the condition $R(T,x_b)=R_2$ post-bounce (see Fig. \ref{fig_shells}), the coordinate $v$ diverges ($v \to \infty$) while $u$ remains finite. Indeed, $T_\mathrm{Cauchy}$ marks the time at which the dust ball surface crosses the Cauchy horizon (see Fig. \ref{fig:asymmetric_collapse}).

Since we have thus far obtained the solution $\tilde{\mathcal{T}}(T,x_b)$ only up to $T_\mathrm{Cauchy}$, we must now perform a gluing with the external vacuum for times beyond this point. To find $\tilde{\mathcal{T}}(T,x_b)$, we again solve the ODE system \eqref{dyn_junction1} and \eqref{dyn_junction2} using the same methods as in Sec. \ref{sec:analysis_of_asymmetric}, but now with the initial conditions $T^\mathrm{bs}_2$ and $\tilde{\mathdutchcal{x}}(T^\mathrm{bs}_2)$. This yields the solution in the interval $T \in (T_\mathrm{Cauchy}, T_3)$, where $T_3$ is defined by the post-bounce condition $R(T_3,x_b) = R_3$. For the results in Fig. \ref{fig:conformal_vacuum}, we have $v \to -\infty$ as $T \to T_3$, with $u(T_3)$ remaining finite. Analogously, for the remaining interval $T \in (T_3, \infty)$, we solve the ODE system \eqref{dyn_junction1} and \eqref{dyn_junction2} with initial conditions $T^\mathrm{bs}_3$ and $\tilde{\mathdutchcal{x}}(T^\mathrm{bs}_3)$. Similarly, for Fig. \ref{fig:asymmetric_collapse}, $v \to \infty$ as $T \to \infty$, whereas surprisingly the numerical results indicate that $u(\infty)$ remains finite. This last property will be discussed later in more detail. We emphasize that when crossing the junction (gluing) surfaces, the vacuum remains in the $\sigma = 1$ dynamical regime; accordingly, we set $\sigma = 1$ for all $T$ in Eqs. \eqref{null_traj}.

To compactify the double-null coordinates introduced above for the exterior vacuum, we define
\begin{equation} \label{uv_compact}
\begin{aligned}
 \tilde{u}(u)&= \pm 1 / \pi \tan^{-1} \left(\kappa u\right)+c_u, \\
 \tilde{v}(v)&= \pm 1 / \pi \tan^{-1} \left(\kappa v \right)+c_v .
\end{aligned}
\end{equation}
where $c_u, c_v \in \{0, 1, 2, 3\}$ and $\kappa$ is a dimensionless parameter introduced to control a visual appearance of the conformal diagram. The $c_u, c_v$ are conformal diagram block constants \cite{Bobula:2024chr}, introduced to stitch all local $(u,v)$ patches into a single global coordinate chart $(\tilde{u}, \tilde{v})$. Additionally, the $\pm$ signs guarantee the monotonicity of the coordinate chart, as illustrated by the following example. Tracing the trajectory of the dust ball surface near $R_3$ before the bounce, we find $u \to \infty$ when approaching from the past, and $u \to -\infty$ when approaching from the future. Therefore, to maintain monotonicity across $R_3$, we select the minus sign for the $\tilde{u}$ coordinate and update the block constant from $c_u=0$ to $c_u=1$ after the crossing. Indeed, the signs and block constants for either $\tilde{u}$ or $\tilde{v}$ must be switched whenever crossing a null horizon where the corresponding $u$ or $v$ diverges. We refer the reader to our earlier works \cite{Bobula:2024chr, Bobula:2024jlh} for a more detailed explanation of this compactification procedure.

To cover the cosmological interior of the asymmetric OS collapse scenario, we first employ standard double null coordinates for the line element \eqref{frw}. Based on this, we show how to extend the exterior global coordinate chart \eqref{uv_compact} to cover the interior as well. We write the null coordinates for the interior as
\begin{equation}
    \begin{aligned}
        u_\mathrm{in}  &= \int^T_0 \frac{1}{a(\tilde{T})} \dd \tilde{T} - x \,, \\
        v_\mathrm{in}  &= \int^T_0 \frac{1}{a(\tilde{T})} \dd \tilde{T} + x  \,.
    \end{aligned}
\end{equation}
Then the extension of \eqref{uv_compact} to cover the interior is given by 
\begin{equation}
\begin{aligned}
& \tilde{u}(T, x)= \pm \tan^{-1} \left[\kappa u\left(T\left\{u_{\text {in }}(T, x)\right\}\right)\right] / \pi+c_u \,, \\
& \tilde{v}(T, x)= \pm \tan^{-1} \left[\kappa v\left(T\left\{v_{\text {in }}(T, x)\right\}\right)\right] / \pi+c_v \,,
\end{aligned}
\end{equation}
where $T\left\{u_{\text{in}}\right\}$ and $T\left\{v_{\text{in}}\right\}$ are, respectively, the inverses of $u_{\mathrm{in }}\left(T, x_b\right)$ and $v_{\mathrm{in }}\left(T, x_b\right)$. For a general discussion of the extension presented here, see \cite{Bobula:2024ywp}. 

Having detailed each step in the construction of the conformal diagram, we now turn to the results presented in Fig.~\ref{fig:asymmetric_collapse}. In the spacetime region inside and near the dust ball prior to the bounce, the causal structure is qualitatively the same as in the symmetric OS collapse scenario (see, e.g., Fig.~1 in Ref.~\cite{Bobula:2024chr}). Specifically, the collapsing dust ball surface consecutively crosses a pair of null horizons: outer $R_3$, and then inner one $R_2$. Within the interior, the trapping horizon connects these null horizons. After the bounce, the causal structure becomes more complex. As explained in Sec.~\ref{sec:analysis_of_asymmetric}, the dust ball surface becomes a $C^0$ shock surface. This shock-like character manifests in the following way: a second trapping horizon emerges at the dust ball surface at radius $R_1$ and traverses the interior. As numerical analysis indicates, it ultimately terminates at the interior spacelike infinity as the interior time $T\to \infty$. Crucially, this trapping horizon is not joined by an exterior null horizon. Instead, shortly afterward, the null horizon $R_2$ joins the surface of the dust ball, and this horizon has no trapping horizon extension into the interior. There is a Cauchy horizon present in the spacetime located at $R_2$ and then extending into the interior of dust ball (extension not plotted on the diagram). Furthermore, numerical analysis indicates that the dust ball surface reaches future null infinity, in contrast to approaching future timelike infinity as in the symmetric OS collapse scenario \cite{Bobula:2024chr}. This is not surprising, as the vacuum exterior to this surface is described by power-law dynamical phase, whereas the dust ball expands exponentially. Alternatively, one can view this vacuum region—characterized by null horizons at $R_2$ and $R_3$—as the domain where $\eta > \eta_\mathrm{min}$. In contrast, within the matched interior after the bounce, $\eta < \eta_\mathrm{min}$. Because the shock follows a non-geodesic path in the vacuum domain $\eta < \eta_\mathrm{min}$, this everywhere-timelike trajectory can asymptotically reach $\mathcal{J}^+$, much like an accelerated observer in Minkowski spacetime.

In the vacuum region beyond the minimal radius $R_\mathrm{min}$, the causal structure resembles that of de Sitter spacetime. To plot this region, we set $\sigma=-1$ in \eqref{uv_compact}. There are null horizons at the radius $R_1$, and there is a spacelike infinity as $R\to \infty$. For this vacuum region, which is bounded by the surface $R_\mathrm{min}$, the condition $\eta < \eta_\mathrm{min}$ holds at every spacetime point. The spacetime in Fig. \ref{fig:asymmetric_collapse} can be further analytically extended beyond the crossing pair of $R_1$ horizons.

\section{Vacuum solution(s)} \label{sec:vacuum}

In GR the purely vacuum Schwarzschild spacetime serves as reference against which the dynamical geometry describing a spherically symmetric matter collapse and black hole formation is compared. Similarly, in our studies it is useful to consider purely vacuum solutions to the loop quantum-corrected LTB dynamics presented in Sec. \ref{sec:asymmetric_review} to provide an analogous reference. In this section we study such solutions. Specifically, we explicitly derive two distinct vacuum solutions in Schwarzschild-like coordinates, analyze their global causal structure, and demonstrate how these two vacua join together. The existence of these two solutions is a consequence of the two dynamical regimes in the OS collapse scenario: power-law contraction and exponential expansion. These results are summarized by the conformal diagram for the purely vacuum spacetime in Fig. \ref{fig:conformal_vacuum}.

The two vacuum solutions are obtained from Eq. \eqref{Rsol} by setting the mass to a constant. These can be expressed in Painlev\'e-Gullstrand-like coordinates as in Eq. \eqref{PGmet}, with the shift function given by Eq. \eqref{shift}. However, in order to relate the results of our studies with those regarding the symmetric collapse, understand the origin of the shock as well as to understand the classical limit of investigated scenario it is useful to switch to Schwarzschild-like coordinates. In order to transform to these coordinates, we introduce the differential
\begin{equation}
\mathrm{d}t
:=
\mathrm{d}T
-
\frac{N}{1-N^2}\,\mathrm{d}R
\end{equation}
which satisfies the necessary condition \cite{Bobula:2024chr} for the existence of the corresponding coordinate $t$, namely $\mathrm{d}(\mathrm{d}t) = 0$. Under this transformation, the metric \eqref{PGmet} takes the form
\begin{equation} \label{met_Schwarz}
\mathrm{d}s^2
=
-
f(R)
\mathrm{d}t^2
+
\frac{\mathrm{d}R^2}{
f(R)
}
+
R^2 \mathrm{d}\Omega^2
\end{equation}
where
\begin{widetext}
\begin{equation} \label{fR}
 f(R):=   1-N^2= 1 - \frac{2^{4/3} G M \left[ -4 \alpha_\Delta^2 G \gamma^2 M + R^3 + \Xi(R) \sigma \right] \left[ -8 \alpha_\Delta^2 G \gamma^2 (1 + \gamma^2) M + R^3 + \Xi(R) \sigma \right]^{4/3}}{\left( R^3 + \Xi(R) \sigma \right)^{8/3}}
\end{equation}
and the value of the parameter $\sigma$ selects the specific vacuum solution. Expanding Eq. \eqref{fR} for large $R$ (that is, $R \to \infty$), we obtain
\begin{equation} \label{fR_expanded}
\begin{aligned}
    f_\text{Sch}(R) &:= f(R)\Big|_{\sigma=1} \\ 
    &= 1 - \frac{2GM}{R} + \frac{4\alpha_\Delta^2 \gamma^2 G^2 M^2 (4 + 3\gamma^2)}{R^4} + \frac{16\alpha_\Delta^4 \gamma^6 G^3 M^3 (1 + \gamma^2)}{R^7} + \frac{48\alpha_\Delta^6 \gamma^8 G^4 M^4 (1 + \gamma^2)^2}{R^{10}} + O\left(\frac{1}{R^{13}}\right) \\[1ex]
    f_\text{dS}(R) &:= f(R)\Big|_{\sigma=-1} \\
    &= 1-\frac{R^2}{\alpha_\Delta^2 (1+\gamma^2)^2}  + \frac{2GM(5\gamma^2 - 1)}{(1+\gamma^2)R} + \frac{4\alpha_\Delta^2 G^2 \gamma^2 (4 - 3\gamma^2)M^2}{R^4} - \frac{16\alpha_\Delta^4 G^3 \gamma^6 (1+\gamma^2)M^3}{R^7} + O\left(\frac{1}{R^8}\right) \, .
\end{aligned}
\end{equation}
\end{widetext}
From this expansion, one sees that in the $\sigma=1$ domain the first two terms of $f_\mathrm{Sch}(R)$ coincide with $f(R)$ corresponding to Schwarzschild black hole geometry. Consequently, the vacuum described by $f_\mathrm{Sch}(R)$ can be interpreted as corresponding to a quantum-corrected Schwarzschild-like black hole geometry, featuring infinitely many corrections weighted by consecutive powers of $\alpha_\Delta$. Interestingly, in the symmetric OS collapse scenario, this series appears to truncate at the term proportional to $1/R^4$, as seen in the vacuum solution derived in Refs. \cite{Kelly:2020lec, Kelly:2020uwj, Husain:2022gwp, Giesel:2022rxi, Bobula:2023kbo, Lewandowski:2022zce, Bobula:2024chr}. 
Similarly, for the vacuum solution in the $\sigma = -1$ domain, the first two terms of $f_\mathrm{dS}(R)$ coincide with $f(R)$ representing de Sitter spacetime. Thus, $f_\mathrm{dS}(R)$, can be treated as describing a modified de Sitter spacetime, also featuring an infinite series of quantum corrections. For the solution given in Eq. \eqref{Rsol}, the Schwarzschild-like and de Sitter-like vacua correspond to the $\eta > \eta_\mathrm{min}$ and $\eta < \eta_\mathrm{min}$ domains, respectively (see also the behavior of $R(\eta)$ in Fig. \ref{fig:Reta}). It is worth noting that metric functions similar to those in Eqs. \eqref{fR} were also derived in Ref. \cite{Ou:2025bbv} by matching an asymmetric dust loop quantum cosmological spacetime with an undetermined spherically symmetric vacuum.

\begin{figure}
    \centering
    \includegraphics[width=1\linewidth]{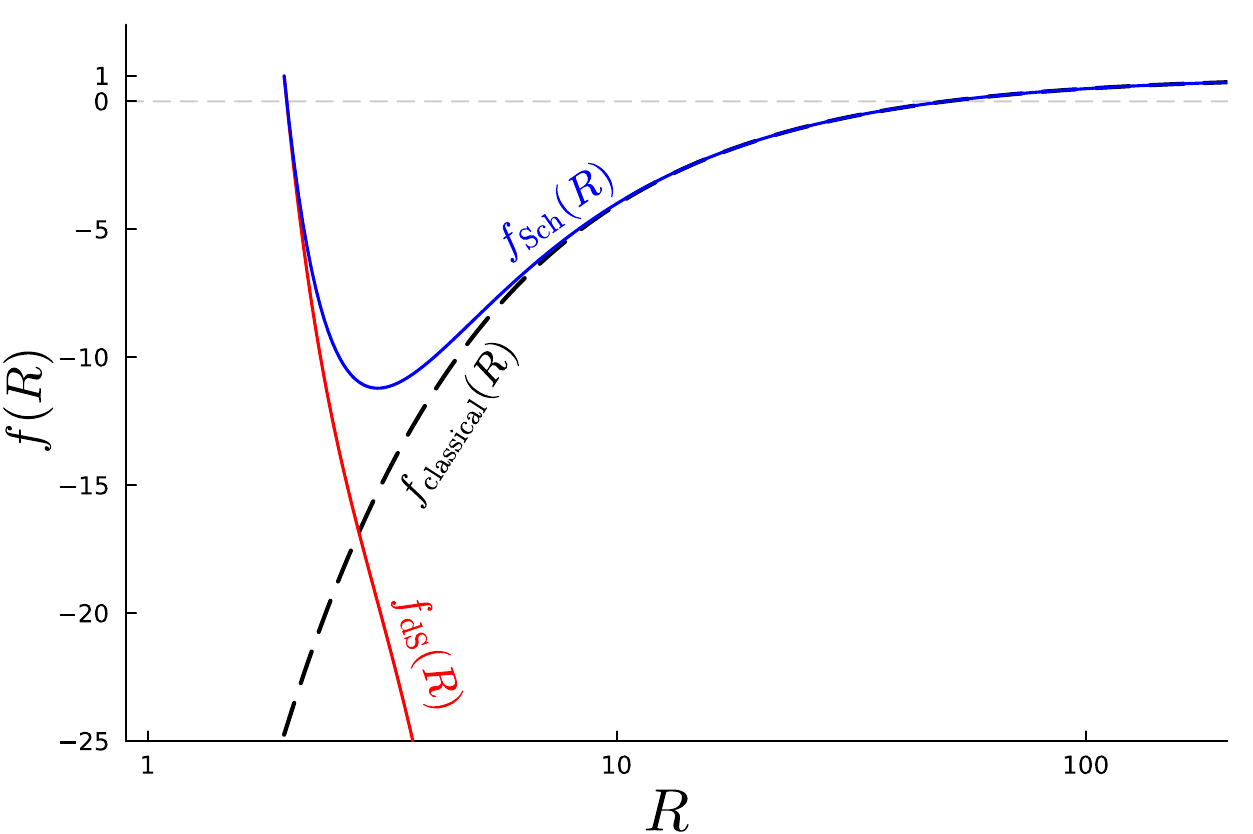}
    \caption{Metric functions $f_\mathrm{Sch}(R)$ and $f_\mathrm{dS}(R)$ (solid curves) for the line element \eqref{met_Schwarz}, representing the quantum-corrected Schwarzschild-like and de Sitter-like vacua. The classical Schwarzschild metric function (dashed curve) is shown for comparison. The parameters are identical to those used in Fig. \ref{fig_shells}, yielding a total mass $M \approx 25$ in Planck units. }
    \label{fig:fschwarz}
\end{figure}

\begin{figure*}
    \centering
    \includegraphics[width=0.75\linewidth]{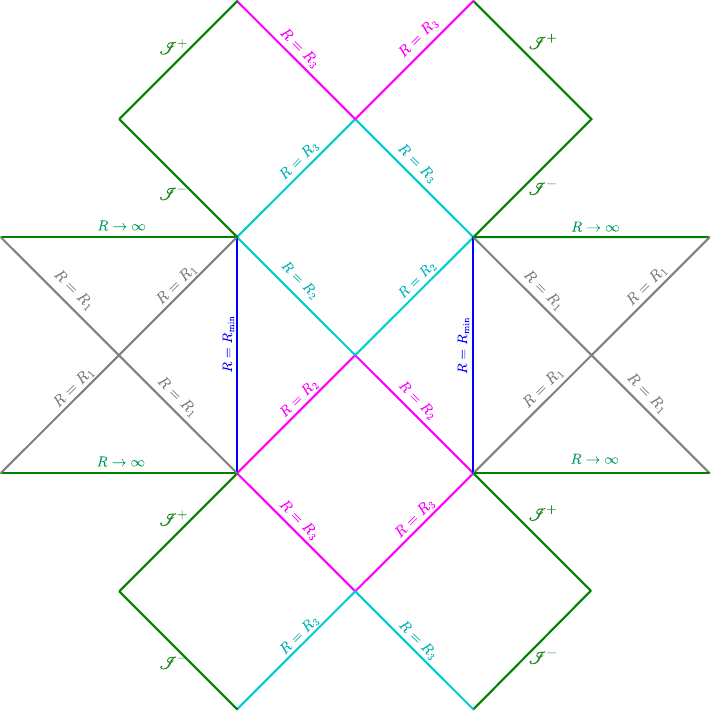}
    \caption{Schematic conformal diagram for the pure vacuum of the studied dynamics. The central vacuum region containing the $R_2$ and $R_3$ horizons is governed by $f_\mathrm{Sch}(R)$ (bounded by $R_\mathrm{min}$), while the regions beyond $R_\mathrm{min}$ are described by $f_\mathrm{dS}(R)$ with horizons at $R_1$. Null and spacelike infinities are indicated by green lines and $R \to \infty$, respectively. The diagram admits analytical extensions vertically beyond $R_3$ and horizontally beyond $R_1$ to generate an infinite grid of identical copies. }
    \label{fig:conformal_vacuum}
\end{figure*}

We plot the metric functions $f_\mathrm{Sch}(R)$ and $f_\mathrm{dS}(R)$ in Fig. \ref{fig:fschwarz}. For comparison, we also include the classical Schwarzschild function, $f_\mathrm{classical}(R) := 1 - \frac{2GM}{R}$. The quantum-corrected function $f_\mathrm{Sch}(R)$ exhibits two roots, corresponding to the horizons at $R_2$ and $R_3$, whereas $f_\mathrm{dS}(R)$ has only a single root at $R_1$. Both functions originate at $R_\mathrm{min}$. While it is natural to ask whether these solutions can be extended to radii smaller than $R_\mathrm{min}$, we find that doing so renders both $f_\mathrm{Sch}(R)$ and $f_\mathrm{dS}(R)$ complex-valued. As an aside, this observation further supports the validity of the extension beyond $R_\mathrm{min}$ presented in Sec. \ref{sec_sub_beyond_min}. Indeed, once $R_\mathrm{min}$ is approached from above along a null geodesic, a transition between the vacua accompanied by the subsequent growth of $R$ appears to be the only viable option.

Based on analysis in Sec. \ref{sec_sub_beyond_min}, one can see that $\eta$ serves as a globally monotonic, albeit non-affine, parameter along radial null geodesics in the vacuum. Along every radial null geodesic, the function $R(\eta)$ shown in Fig. \ref{fig:Reta} is fully explored as the affine parameter varies from $-\infty$ to $\infty$, accompanied by a switch from $R$ to $-R$ (up to additive and multiplicative constants). That is, for each geodesic, all values of $\eta \in (\eta_\mathrm{inf}, \infty)$ are covered, with the areal radius spanning $R \in (R_\mathrm{min}, \infty)$ and the affine parameter ranging from $-\infty$ to $\infty$. This unambiguously determines the the causal structure in these regions. In consequence, we now have sufficient data to construct the conformal diagram for the purely vacuum solution of the studied dynamics, which is presented in Fig. \ref{fig:conformal_vacuum}. In this diagram, we see that regions described by $f_\mathrm{Sch}(R)$ and $f_\mathrm{dS}(R)$ join each other along surfaces $R_\mathrm{min}$. 

For $f_\mathrm{Sch}(R)$, radial timelike geodesics behave analogously to those in symmetric OS collapse. They originate at past timelike infinity (the intersection of $R_3$ and $\mathcal{J}^-$), cross $R_3$ and the $R_2$ horizon, bounce at $R_\mathrm{min}$, and recross $R_2$ and $R_3$, ultimately reaching future timelike infinity (the intersection of $R_3$ and $\mathcal{J}^+$). Similarly, for $f_\mathrm{dS}(R)$, radial timelike geodesics originate at past spacelike infinity, cross $R_1$, bounce at $R_\mathrm{min}$, recross $R_1$, and terminate at future spacelike infinity.

The vacuum conformal diagram in Fig. \ref{fig:conformal_vacuum} is related to the collapse diagram in Fig. \ref{fig:asymmetric_collapse} as follows. The surface of the dust ball in Fig. \ref{fig:asymmetric_collapse} slices through the vacuum in Fig. \ref{fig:conformal_vacuum}—starting near the past null infinity at the bottom right and terminating at the future null infinity at the top right. Consequently, essentially only the right-hand vacuum sector persists, which is then matched across the boundary to the interior geometry of the dust ball.

We emphasize that, as established in prior work \cite{Giesel:2024mps}, the Kretschmann scalar diverges at the surface $R_\mathrm{min}$. This surface is present in the pure vacuum conformal diagram in Fig. \ref{fig:conformal_vacuum}, as well as in the vacuum exterior of the dust ball in Fig. \ref{fig:asymmetric_collapse}. However, based on the analysis presented in this work, we interpret this divergence as a shock surface arising from a mismatch between the de Sitter-like and Schwarzschild-like dynamical domains. Specifically, we do not expect the de Sitter-like and Schwarzschild-like vacuum regions to join smoothly ($C^1$ continuous) across this surface. This behavior is in direct analogy to a well-known classical result: the Schwarzschild solution cannot be smoothly matched to a de Sitter spacetime \cite{Poisson:1988wc}. In the present model, we expect that this matching can only be achieved at the cost of discontinuous extrinsic curvatures, which explains the origin of the diverging Kretschmann scalar.

%%%%%%%%%%%%%%%%%%%%%%%%%%%%%%%%%%%%%%%%%%%%%%%%%%%%%%%%%%%%%%%%%%

%%%%%%%%%%%%%%%%%%%%%%%%%%%%%%%%%%%%%%%%%%%%%%%%%%%%%%%%%%%%%%%%%%
\section{Discussion} \label{sec:discussion}

We investigated asymmetric OS collapse within the framework of symmetry-reduced LQG. Specifically, we incorporated the effective dynamics of a recently generalized cosmological asymmetric bounce from LQC \cite{Assanioussi:2018hee, Assanioussi:2019iye} into LTB spacetimes \cite{Giesel:2024mps, Fazzini:2026uqp, Giesel:2026pjj}. This allowed us to analyze the fully dynamical collapse of a homogeneous dust ball into a loop quantum-corrected black hole.

The key question we answered in this work is how asymmetric bounce dynamics modify the physical picture of the mainstream symmetric bounce provided by LQC. In this context, our main result—summarized by the numerically computed conformal diagram of the asymmetric OS collapse scenario in Fig. \ref{fig:asymmetric_collapse}—demonstrates that the resulting spacetime is geodesically complete. Specifically, the dust ball collapses following a power law, crosses the inner and outer black hole horizons, and then violently reexpands in an exponential regime. It then crosses corresponding horizons once more before reemerging into a new asymptotic region. During this violent reexpansion, we found that a $C^0$ shock is present at the surface of the dust ball—that is, a discontinuity in the extrinsic curvature while the induced metric remains continuous. The presence of this shock is the key difference compared to the symmetric collapse \cite{Kelly:2020lec, Kelly:2020uwj, Husain:2021ojz, Giesel:2022rxi, Husain:2022gwp, Lewandowski:2022zce, Bobula:2023kbo, Fazzini:2023scu, Bobula:2024chr}. This shock originates from a dynamical mismatch: the interior of the dust ball has already entered the exponential expansion regime, while the exterior vacuum shells (timelike geodesics) impacting the surface are still in the power-law contraction regime. 

A crucial advantage of the asymmetric collapse is the absence of the timelike singularity present in the symmetric case. In the symmetric scenario, an analytic extension of the vacuum solution leads to a Reissner-Nordström-like singularity located beyond the Cauchy horizon \cite{Bobula:2023kbo, Lewandowski:2022zce, Bobula:2024chr}. In our detailed study of the vacuum sector for the asymmetric OS collapse, we found that it cannot be written using a single metric in Schwarzschild-like coordinates. Instead, two metrics are necessary to fully describe the vacuum. The need for these two metrics originates from two distinct dynamical domains: the power-law regime and the exponential regime. These regimes yield a quantum-corrected Schwarzchild-like black hole metric and a corrected de Sitter-like metric, respectively, both given in Eqs. \eqref{fR_expanded}. These vacua join each other at the minimal areal radius $R_\mathrm{min}$ in a $C^0$ manner (where the discontinuity in the extrinsic curvature is signaled by the divergence of the Kretschmann scalar). Despite this divergence, we found that geodesics—particularly null ones—are extendible beyond this surface in a way that resolve the singularity problem of the symmetric case. Specifically, the areal radius along a null geodesic changes its monotonicity after crossing $R_\mathrm{min}$. This reversal provides geodesic completeness and yields a rich causal structure, summarized in Figs. \ref{fig:asymmetric_collapse} and \ref{fig:conformal_vacuum} for the dust ball collapse and pure vacuum cases, respectively.

This work demonstrates that a strict implementation of Thiemann's regularization \cite{Thiemann:1996av, Thiemann:1996aw}, yields a more consistent (and nonsingular) dynamics than the mainstream LQC based prescriptions. The asymmetric bounce dynamics is often regarded as being closer to full LQG, as it starts from the Hamiltonian constraint of the full theory on an arbitrary spatial background, retaining the Euclidean and Lorentzian parts of the constraint as separate objects throughout the quantization procedure and regularizing them via separate identities holding without symmetry assumptions.
In contrast, mainstream LQC \cite{Ashtekar:2006rx, Ashtekar:2006uz, Ashtekar:2006wn} splits the Lorentzian part of the Hamiltonian constraint onto sum of a term proportional to the Euclidean part and $3$-dimensional Ricci scalar. The latter (though its quantization within LQG framework is now available \cite{Alesci:2014aza}) is then neglected as for the considered spacetimes it classically vanishes. This simplification leads to the familiar symmetric bounce. 
In this work, we demonstrated how these different quantization strategies manifest in black hole solutions. In the metric derived in Eq. \eqref{fR_expanded}, the metric function can be viewed as a classical part supplemented by infinitely many quantum corrections. Each correction is weighted by powers of $\alpha_\Delta$, which provides the characteristic length scale for this dynamical framework. In the symmetric case, however, such corrections truncate at the term proportional to $1/R^4$ for the quantum-corrected Schwarzschild-like black hole.

We argue that our proposed approach to analyzing the $C^0$ shock is not restricted to frameworks rooted in LQG, but has wider applicability. Specifically, we believe that such shocks are highly likely to occur in any dynamical theory that reproduces shell trajectories on the areal radius versus proper time plane, similar to those shown in Fig. \ref{fig_shells}. Such a plot fully encodes the spacetime geometry, as it determines the metric functions entering the LTB metric—namely, $R(T,x)$ and its derivative with respect to $x$. In our analysis of the shock, we assumed only: (i) the consistency of the tangent vectors through coordinate transformations, as seen from both inside and outside the junction surface, and (ii) the continuity of the areal radius at the junction surface. These minimal and, we believe, theory-agnostic geometric requirements allow to fully characterize the shock. However, one limitation of our current analysis is the assumption that the junction surface coincides with the surface of the collapsing dust ball, a restriction that could be relaxed in general.

An open problem we leave for future investigation is verifying whether a situation similar to classical GR occurs in the studied model — namely, that the discontinuity in the extrinsic curvature signals the presence of a distributional form of matter, such as thin shells \cite{Poisson:2009pwt}. Timelike thin shells were studied in detail for symmetric collapse dynamics with inhomogeneous initial dust profiles, and they were shown to be viable extensions beyond shell-crossing singularities \cite{Bobula:2026zlq}. The fate of such thin shells in the asymmetric but homogeneous collapse case remains open, along with the question of whether their potentially non-zero masses could spark dynamical inconsistencies in the physical picture. This analysis requires studying the field equations in a distributional formulation, similar to the developments provided in Refs. \cite{ Fazzini:2025nse, Bobula:2026zlq, Sahlmann:2025fde, Qu:2026vvs} for the case of loop quantum symmetric collapse.

This work proves the first fully dynamical formation of a regular black hole within a framework based on LQC, successfully reproducing the cosmological dynamics in the homogeneous and isotropic reduction. The studied dynamics contributes to the broader spectrum of regular black hole formation scenarios recently proposed within the literature such as those based/inspired-by asymptotic safety \cite{Bonanno:2023rzk} or string theory \cite{Bueno:2024eig}. It should be emphasized that earlier models of regular black hole formation generally rely on exotic matter \cite{Shojai:2022pdq}, ad hoc degrees of freedom, or highly fine-tuned fields such as nonlinear electrodynamics \cite{Malafarina:2022oka}. However, one could object that the black hole presented in this work is not fully regular due to the presence of shocks and the divergence of the Kretschmann scalar at $R_\mathrm{min}$. In response to such objections, we emphasize that the studied geometry is geodesically complete. The aforementioned non-smoothness does not represent a breakdown in the predictability of the theory — unlike the central or shell-crossing singularities in classical GR (where no unambiguous extension has been constructed so far). Instead, the shocks here are fully controlled, originating from a dynamical process characterized by two distinct phases.

Finally, another open challenge recurring in the literature on regular black hole formation is the presence (formation) and instability of Cauchy horizons (which mark the boundaries of predictability for the dynamical equations) along with the related mass inflation instability \cite{Poisson:1989zz, Poisson:1990eh, Ori:1991zz, Liu:2026ltw}. Although models of regular black holes without Cauchy horizons have been proposed to circumvent these issues \cite{Simpson:2018tsi, Carballo-Rubio:2018jzw, Carballo-Rubio:2022kad, Ovalle:2023vvu, Zhang:2024ney, Calza:2025mrt}, to our knowledge, such constructions are not accompanied by a consistent stellar collapse or matching cosmological dynamics capable of successfully modeling the early universe. This recurring pattern suggests that simply eliminating Cauchy horizons in black hole vacuums is not enough.

A possible solution to the Cauchy horizon problem that we find particularly attractive is the regular black hole formation and evaporation scenario proposed by Hayward \cite{Hayward:2005gi}. In this framework, backreaction from Hawking radiation prevents trapping horizons from remaining null throughout the dynamics, thereby eliminating standard event and Cauchy horizons. Such spacetimes can successfully reproduce the phenomenology of classical GR black holes \cite{Bambi:2025wjx} and provide a convincing resolution to long-standing puzzles like the information paradox \cite{Hayward:2005gi, Ashtekar:2005cj}. Nevertheless, recent studies show that even in models where the trapping horizon does not remain null forever and exhibits Hayward's typical closed trajectory, mass inflation can still persist \cite{Carballo-Rubio:2024dca}. This fact motivates a detailed study of backreaction—a topic that remains largely underexplored in the literature.

\appendix

\begin{widetext}
\section{$s(x)$ function} \label{appendix_sx}
The Eqs. \eqref{Rsol}, \eqref{rho0}, \eqref{R0} and \eqref{Mx} together allow us to find $\eta$ on the initial time slice. We choose the positive roots when algebraically solving for $\eta$ at that time, as they correspond to solutions that collapse under forward evolution in $T$. We obtain
\begin{equation} \label{eta0}
\eta(0,x) =
\begin{cases}
\dfrac{1}{6 (\pi \rho_0)^{1/2}}
\left(
3 - 16 \alpha_\Delta^2 \gamma^2 \pi \rho_0
\pm
3 \left(
1 - \dfrac{32 \alpha_\Delta^2 \gamma^2 (1+\gamma^2)\pi\rho_0}{3}
\right)^{1/2}
\right)^{1/2} \,, \;\;
 \text{for } x \le x_b \,, \\[2ex]
\dfrac{1}{6 \pi^{1/2}}
\left(
\dfrac{
3x^3
-16 \alpha_\Delta^2 \gamma^2 \pi x_b^3 \rho_0
\pm
\left(
-96 \alpha_\Delta^2 \gamma^2 (1+\gamma^2)\pi x_b^3 \rho_0 x^3
+9x^6
\right)^{1/2}
}{
x_b^3 \rho_0
}
\right)^{1/2}\,, \;\;
 \text{for } x>x_b\,.
\end{cases}
\end{equation}
We further restrict to the $+$ sign branch in Eq.~\eqref{eta0} to ensure that the resulting expressions remain real-valued. Substituting Eq.~\eqref{eta0} into the expression for $R$ in Eq.~\eqref{Rsol} then yields the desired function
\begin{equation} \label{sx_formula}
s(x)= \eta(0,x)-\frac{2}{3} \alpha_{\Delta}\left(\gamma^2+1\right)  \tanh ^{-1} \left(\frac{2 \alpha_{\Delta} \gamma^2}{3 \eta(0,x)}\right)
\end{equation}

\end{widetext}

% Bibliography linked to sample.bib
\bibliography{sample}

\end{document}